\documentclass[twocolumn,table,openany]{aastex631}
\usepackage{tabularx} 
\usepackage{amsmath}  
\usepackage{amssymb}
\usepackage{dcolumn}
\usepackage{booktabs}

\newcommand{\orcid}[1]{\href{https://orcid.org/#1}{\textcolor[HTML]{A6CE39}{\aiOrcid}}}

\newcommand{\msun}{M$_{\odot}$}
\newcommand{\msundyn}{\ifmmode \mathrm{M}_{\odot}\elseM$_{\odot}$\fi}
\newcommand{\rsun}{\ifmmode \mathrm{R}_{\odot}\elseR$_{\odot}$\fi}
\newcommand{\degree}{\ifmmode^{\circ}\else$^{\circ}$\fi}
\newcommand{\amin}{\ifmmode^{\prime}\else$^{\prime}$\fi}
\newcommand{\asec}{\ifmmode^{\prime\prime}\else$^{\prime\prime}$\fi}
\newcolumntype{d}[1]{D{.}{.}{#1}} 
\begin{document}

\title{A 34 yr Timing Solution of the Redback Millisecond Pulsar Terzan 5A}

\correspondingauthor{Alexandra C. Rosenthal}
\email{acr244@cornell.edu}

\author[0009-0001-2223-2975]{Alexandra C.~Rosenthal}
\affiliation{University of Virginia, Department of Astronomy, Charlottesville, VA 22904, USA}
\affiliation{National Radio Astronomy Observatory, 520 Edgemont Rd., Charlottesville, VA 22903, USA}

\author[0000-0001-5799-9714]{Scott M.~Ransom}
\affiliation{National Radio Astronomy Observatory, 520 Edgemont Rd., Charlottesville, VA 22903, USA}
\affiliation{University of Virginia, Department of Astronomy, Charlottesville, VA 22904, USA}

\author[0000-0002-2764-7248]{Kyle A.~Corcoran}
\affiliation{University of Virginia, Department of Astronomy, Charlottesville, VA 22904, USA}

\author[0000-0002-2185-1790]{Megan E.~DeCesar}
\affiliation{George Mason University, Fairfax, VA 22030, resident at the U.S. Naval Research Laboratory, Washington, D.C. 20375, USA}
\author[0000-0003-1307-9435]{Paulo C.~C.~Freire}
\affiliation{Max-Planck-Institut f\"ur Radioastronomie, Auf dem H\"ugel 69, D-53121 Bonn, Germany}

\author[0000-0003-2317-1446]{Jason W.~T.~Hessels}
\affiliation{ASTRON, the Netherlands Institute for Radio Astronomy, Oude Hoogeveensedijk 4, 7991 PD Dwingeloo, The Netherlands}
\affiliation{Anton Pannekoek Institute for Astronomy, University of Amsterdam, Science Park 904, 1098 XH, Amsterdam, The Netherlands}

\author[0000-0001-5567-5492]{Michael J.~Keith}
\affiliation{Jodrell Bank Centre for Astrophysics, Department of Physics and Astronomy, University of Manchester, Manchester M13 9PL, UK}

\author[0000-0001-5229-7430]{Ryan S.~Lynch}
\affiliation{Green Bank Observatory, PO Box 2, Green Bank, WV 24494, USA}

\author[0000-0002-4799-1281]{Andrew Lyne}
\affiliation{Jodrell Bank Centre for Astrophysics, Department of Physics and Astronomy, University of Manchester, Manchester M13 9PL, UK}

\author[0000-0002-6709-2566]{David J. Nice}
\affiliation{Department of Physics, Lafayette College, Easton, PA, 18042, USA}

\author[0000-0001-9784-8670]{Ingrid H.~Stairs}
\affiliation{Dept.~of Physics and Astronomy, UBC, 6224 Agricultural Road, Vancouver, BC V6T 1Z1 Canada}

\author[0000-0001-9242-7041]{Ben Stappers}
\affiliation{Jodrell Bank Centre for Astrophysics, Department of Physics and Astronomy, University of Manchester, Manchester M13 9PL, UK}

\author[0000-0002-1468-9668]{Jay Strader}
\affiliation{Center for Data Intensive and Time Domain Astronomy, Department of Physics and Astronomy,\\ Michigan State University, East Lansing, MI 48824, USA}

\author[0000-0002-2025-9613]{Stephen E. Thorsett}
\affiliation{Department of Physics, Willamette University, Salem, OR 97301, USA}

\author[0000-0003-1814-8620]{Ryan Urquhart}
\affiliation{Center for Data Intensive and Time Domain Astronomy, Department of Physics and Astronomy, Michigan State University, East Lansing, MI 48824, USA}

\received{2024 October 28}
\revised{2025 January 27}
\accepted{2025 January 27}
\published{2025 March 27}
\reportnum{ApJ 2025, v892, n2}


\begin{abstract}
    We present a 34-year timing solution of the redback pulsar system Terzan~5A (Ter5A). Ter5A, also known as B1744$-$24A or J1748$-$2446A, has a 11.56\,ms pulse period, a $\sim$0.1\,\msun\ dwarf companion star, and an orbital period of 1.82\,hours.  Ter5A displays highly variable eclipses and orbital perturbations. Using new timing techniques, we have determined a phase-connected timing solution for this system over 34 years. This is the longest ever published for a redback pulsar. We find that the pulsar's spin variability is much larger than most globular cluster pulsars. In fact, of the nine redback pulsars with published or in preparation long-term timing solutions, Ter5A is by far the noisiest. We see no evidence of strong correlations between orbital and spin variability of the pulsar. We also find that long-term astrometric timing measurements are likely too contaminated by this variability to be usable, and therefore require careful short-term timing to determine reasonable positions. Finally, we measure an orbital period contraction of $-2.5(3) \times 10^{-13}$, which is likely dominated by the general relativistic orbital decay of the system. The effects of the orbital variability due to the redback nature of the pulsar are not needed to explain the observed orbital period derivative, but they are constrained to less than $\sim$30\% of the observed value.
\end{abstract}

\section{Introduction}

The high stellar density and resulting high rate of gravitational encounters in globular clusters make them uniquely productive birthplaces for both millisecond pulsars (MSPs) and other types of exotic compact binaries. Globular clusters produce two orders of magnitude more low-mass X-ray binaries per unit mass than the galactic disk \citep[e.g.][]{XRBGCs1, XRBGCs2}, which results in a similar overproduction of MSPs and exotic systems. To date, 343 pulsars have been identified in 45 globular clusters\footnote{\url{https://www3.mpifr-bonn.mpg.de/staff/pfreire/GCpsr.html}}, and 310 of those are MSPs (pulse period $< 20$\,ms). That number will only increase as pulsar searches with new and more sensitive telescopes continue.

The interactions between pulsars and cluster stars add accelerations to both the pulsar's spin and orbital parameters \citep{Prager2017}. These effects mean that globular cluster pulsars are not useful for certain high-precision timing applications, such as probing the nHz gravitational wave background, but the sheer number of cluster pulsars provides a rich dataset for probing physics such as cluster dynamics \citep{ClustDynProbes}, cluster gas properties \citep{Abbate2018,Prager2017}, general relativity \citep{FreireWex2024}, and the neutron star equation of state \citep{LattimerEOS,OzelFreire}. 

``Spider'' pulsars are one of the exotic and complex types of interacting binaries found in globular clusters. Spider pulsars are a class of MSP with relativistic winds that strongly affect a close companion star. These interesting MSP binaries are so named because ablation from the pulsar wind causes mass-loss indicative of the MSP ``eating’’ its companion. The pulsar's ablation of its companion creates an ionized wind that delays or completely obscures pulses and perturbs the canonical ``perfect'' astrophysical clock. This ablated material may approach the pulsar, or, more likely, be ejected from the system entirely. It is important to note that this mass transfer mechanism is separate from Roche lobe overflow, even though the irradiation from the pulsar likely helps keep the companion inflated to approximately its Roche lobe radius \citep{Sivan2021}. 

One sub-class of spider pulsars are the redbacks. Redback systems have main sequence companions of mass M$_c \gtrsim 0.1$\,\msun \citep{Roberts2013, Chen_2013}. These redback companions are brighter and have larger radii than isolated main-sequence stars of the same spectral type \citep{DeVito2020}. Furthermore, redbacks have much more irregular, and typically longer duration, radio eclipses than other classes of spider pulsars. In the galactic field, such a system forms from a binary with a short orbital period and a main sequence companion that filled its Roche lobe. There are 17 known redback pulsars in globular clusters\footnote{\url{https://www3.mpifr-bonn.mpg.de/staff/pfreire/GCpsr.html}}, and more are likely to be found. Redback formation is not well understood (see \cite{DeVito2020} for a summary of formation paths), but must involve a significant period of mass transfer \citep{Chen_2013}, as redbacks are disproportionately represented among the fastest spinning MSPs, comprising 5 of the 15 fastest-spinning pulsars\footnote{\url{https://www3.mpifr-bonn.mpg.de/staff/pfreire/GCpsr.html}} (and the fastest MSP known is another Ter5 redback, Ter5ad, \citealt{Ter5ad}). Once a system is visible as a redback, mass transfer isn't presently occurring \citep{Roberts2013}. Whether redbacks represent a temporary pause in the transitional MSP state, such as that observed in M28I, J1023$+$0038, and PSR J1227$-$4853 \citep{Papitto_M28I, Stappers_tMSP, Patruno_tMSP,Third_tMSP1,Third_tMSP2} is unclear. 

This paper focuses on Ter5A, which is one of the most compact redback systems known and has the shortest orbital period of any redback. It was the second spider pulsar system discovered, and the first redback system detected \citep{Discovery,Nice1990}, though at the time the distinction between the classes of spider pulsars had not been established. It is the brightest of the 49 confirmed MSPs in Terzan~5 \citep{Martsen2022,Pad49}, and one of the furthest MSPs from the cluster core. The dense cluster introduces higher-order accelerations to the pulsar, which perturb its observed spin rate beyond the standard spindown over time. Ter5A is a uniquely noisy MSP, with timing residuals on the order of milliseconds over decade timescales. This is two orders of magnitude larger than most other known MSPs \citep{Nice2000Proceedings}. Ter5A is also unusually \textit{less} recycled than almost all other spider pulsars: most spider pulsars have spin periods under 10\,ms, while Ter5A's period is 11.56\,ms (a notable exception is the redback B1718-19, which has a 1.004\,s period \citealt{Lyne1993}).

Ter5A's companion is a $\sim$0.1\,\msun\ star in a 1.82 hour compact orbit. The ionized wind blown off the companion eclipses the pulsar, most likely via synchrotron absorption \citep{Polzin2018}. With some assumptions regarding the pulsar's magnetic moment, the companion's mass loss rate due to irradiation by the pulsar is $2 \times 10^{-12}$\,\msun\,yr$^{-1}$ \citep{Shaham1995}, which results in an evaporation timescale of $\sim$50 Gyr. \cite{ThorsettNice1991} also posit a wind model that does not ablate the companion within a Hubble time. \cite{Shaham1995} predicts that this rate is on the threshold between the conditions for accretion of this ablated wind onto the pulsar and expulsion from the system in absence of accretion. In addition to causing eclipses, interactions between the companion and the pulsar produce orbital variations. Ter5A's companion also causes many other interesting physical effects on the propagation of the pulsar signal \citep{Bilous2019BSPs,Li2023}.

The eclipse usually occurs around phase 0.25 of the orbit (i.e. {when the companion is between the pulsar and the observer). However, the eclipsing behavior of the pulsar is highly variable: some observations have pulses even during phase 0.25, and sometimes the pulsar disappears for hours at a time \citep{Discovery,Nice1990}. Additional eclipse-like disappearances at different orbital phases have also been observed \citep{Bilous2019BSPs}. The occasional disappearance is attributed to the system being entirely enshrouded by ablated material. The eclipses are also frequency-dependent: the observing frequency impacts the duration and even the appearance of the eclipse \citep{Nice1990,You2018}. The eclipses are also longer than expected for a companion confined to its Roche lobe, implying some form of continuous mass loss, whether by Roche lobe overflow, stellar wind, or irradiation by the pulsar \citep{Nice2000Proceedings}. Numerical simulations by \cite{TavaniBrookshaw} imply Ter5A's irregular and long-term eclipses can be generated by a companion mass-loss rate of $4 \times 10^{-13}$\,\msun/yr, which is within a factor of ten of that estimated by \cite{Shaham1995}. 

The irregular nature of the eclipse, accelerations from the cluster, and effects from Ter5A’s companion have made long-term timing efforts prohibitively difficult. Accounting for the accelerations is a well-documented process \citep[e.g.][]{Prager2017, Ridolfi_LTRB}, but accounting for the other variations is difficult on long timing baselines because the orbital parameters also change stochastically over time. Without timing methods capable of handling these complications, redback timing has only been attempted on shorter time spans ranging from a few years in length \citep{Archibald2013,Deneva,Prager2017,Miraval,2yr} to a decade \citep{Nice2000Proceedings, Ridolfi_LTRB}. Longer timing baselines have exploited new timing techniques to accommodate orbital variations, such as \cite{Thongmeearkom2024}'s 15-year timing solution, which uses Gaussian Process Regression to track orbital evolution in Fermi data. Additionally, timing the system over shorter time spans has been attempted using custom methods to account for the changing orbital parameters \citep{Nice2000Proceedings,Bilous2011}. 

Using $\sim$20 years of archival Green Bank Telescope (GBT) data, $\sim$10 years of older Very Large Array (VLA) and Green Bank 140-foot telescope data \citep{Nice2000Proceedings}, $\sim$34 years of data from the Lovell Telescope at Jodrell Bank Observatory (JB) dating from the pulsar's 1989 discovery to present, four Parkes Radio Telescope (Parkes) observations, the timing programs \texttt{PINT}\footnote{\url{https://github.com/nanograv/PINT}} \citep{PINT1,PINT2} and \texttt{tempo}\footnote{\url{https://tempo.sourceforge.net/}}, and a new piecewise continuous binary timing model \citep{O'Neill2025}, in addition to a new isolation technique that decouples the orbital variability from the timing of the pulsar, we are able to unambiguously track the pulsar spin evolution over 34 years. This is the longest timing solution for a redback system ever produced, and the longest currently possible. 

This work is organized as follows: In Section~\ref{observationssection}, we detail the observational information for our data. In Section~\ref{timing}, we discuss the methodology used to time this system. In Section~\ref{simplelongterm}, we present a long-term timing model that assumes there are no variations in the pulsar's orbit. In Section~\ref{sfd}, we present a long-term timing solution that accounts for orbital variations. In Sections~\ref{T0var} and \ref{PBDOT}, we analyze the orbital behavior of the system. In Section~\ref{pospm} we discuss the specific challenges in determining the position and proper motion of Ter5A. In Section~\ref{xray} we compare our radio observations to X-Ray observations. In Section~\ref{energetics} we compare our findings to the Applegate (1992) model. In Sections~\ref{torques} and \ref{knownsystematics} we discuss potential and known contaminants in the data, including the possible influence of torques on the pulsar. Finally, in Section~\ref{concs}, we discuss our results.

\section{Observations} \label{observationssection}

Our timing solution incorporates data from five telescopes: The 100\,m Green Bank Telescope (GBT); the Green Bank 140\,ft Telescope (GB140); the Very Large Array (VLA); Murriyang, the Parkes Radio Telescope (Parkes); and the Lovell telescope at Jodrell Bank Observatory (JB).

We started with 244 observations of Ter5A spanning MJDs 47965$-$60144 (i.e.~1990.2$-$2023.5) from GBT, GB140, VLA, and Parkes. The majority of observations were made between MJDs 53193$-$60144 (i.e.~2004.5$-$2023.5) with the GBT primarily with the L-band (1.0-1.8\,GHz) and S-band (1.6-2.5\,GHz) receivers. Over the span of this dataset, the GBT backend processing system changed from SPIGOT \citep{SPIGOT} to GUPPI \citep{GUPPI} to VEGAS \citep{VEGAS}. The SPIGOT observations are described in \cite{Ransom2005}. The GUPPI and VEGAS observations are described in \cite{Martsen2022}. Data prior to MJD 55422 were processed using incoherent dedispersion, and data after 55422 -- most of the GUPPI data and all of the VEGAS data -- were processed using coherent dedispersion. We processed subbanded searchmode data, where the original raw data was partially dedispersed (that is, dedispersed into subbands at a nominal DM) to 238\,pc\,cm$^{-3}$, which is roughly the average DM of the cluster \citep{Ransom2007}, and for the coherently dedispersed data also converted to total intensity and downsampled in time by a factor of four. The data were fully dedispersed into topocentric time series at a DM of 242.34\,pc\,cm$^{-3}$, which is the current DM of the pulsar. We then folded the time series modulo the predicted spin period, and integrated over one-minute intervals in order to produce  pulse times of arrival (TOAs). For the majority of the data, pulse smearing due to the mismatch between subband DM (238\,pc\,cm$^{-3}$) and time series DM (242.34\,pc\,cm$^{-3}$) is $<$40\,$\mu$s, and even in the most affected data it is $<$100\,$\mu$s. There is additional pulse smearing near the early 2000s, as the actual DM at the time was about $\sim$0.2\,pc\,cm$^{-3}$ lower, but the resulting effects ($\sim$130\,$\mu$s smearing for S-band and $\sim$300\,$\mu$s for L-band) are much less than both the pulse width and the residual timing noise.

The pulse template we used for these TOAs was generated from a very strong S-band detection in the SPIGOT data, with the eclipses removed. Because the effect of dispersive smearing is small ($\sim$0.3\,ms relative to Ter5A's 11.56\,ms pulse period and $\sim$1\,ms pulse width), there is no significant impact on our results from using this pulse template across multiple frequency bands \citep[for details, see][]{Ransom2005}. While this would be insufficient for microsecond-precision timing, it is more than sufficient for timing on $\sim$0.1\,ms scale.

Prior to MJD 53193 (August 2004), most of the TOAs are a combination of VLA 1660 MHz data taken with the Princeton Mark3 backend, GB140 800\,MHz and L-band data taken from 1990$-$1999 with the Spectral Processor, and several early GBT observations between 2000$-$2004. Details on these older TOAs can be found in \cite{Nice2000Proceedings,Nice1992}. We incorporate data from Parkes taken in search mode on four days between MJDs 50800$-$52000 at L-band in several different observing modes. We note that the VLA, GB140, and GBT observations prior to MJD 53193 were available only as pre-calculated TOAs and so were generated slightly differently and using different pulse templates from the later data. For the GB140 timing, separate templates were made for each of the two receivers (800 MHz, L-band), in each case by cross-correlating and averaging data from a large number of observations. Those earlier datasets require systematic timing offsets between themselves and the later GBT data (i.e.~\texttt{JUMP}s).

After cleaning and phase-connecting the GBT, GB140, VLA, and Parkes data using the methods outlined in sections \ref{DataCleaning} and \ref{timing}, we compared our results to JB observations spanning from discovery to June 2024. The JB data consist observations of typically between 12 and 30 minutes duration with the Lovell telescope at L-band and were processed by one or more of three backend systems: 1) an incoherently-dedispersing analogue filterbank (AFB) with a total bandwidth of 32 MHz, spanning MJDs 47952-55196 (1990.2-2020.0)  2) an incoherently dedispersing digital fileterbank (DFB) \citep{Ant23} with a total bandwidth of 384 MHz, spanning MJDs 54973-60490 (2009.0-2024.5) and 3) a coherently-dedispersing system based upon a ROACH processor \citep{Ant23} with a total bandwidth of 400 MHz split into two bands, spanning MJDs 55679-60321 (2011.4-2024.0).  The incoherent systems provided TOAs from 3-minute integrations, while the coherent system used 2.5-minute integrations. The JB data was already phase-connected and matched well with our simple timing solution (see Figure \ref{fig:simple}), and our model accommodating changing orbital parameters (see Figure \ref{fig:FreqDerivPlots}), demonstrating that our novel methodology replicates traditional solutions.

Observation information for these eclipse-cleaned data is tabulated in Table~\ref{tab:obssum}. Because these are the cleaned data, the TOA totals in Table~\ref{tab:obssum} differ somewhat from the total set of TOAs accompanying this work. All observations, except those from JB, are band-averaged, and we leave all investigations of short-timescale frequency-dependent timing issues for future work. 

\begin{figure*}
    \centering
    \includegraphics[width=\textwidth]{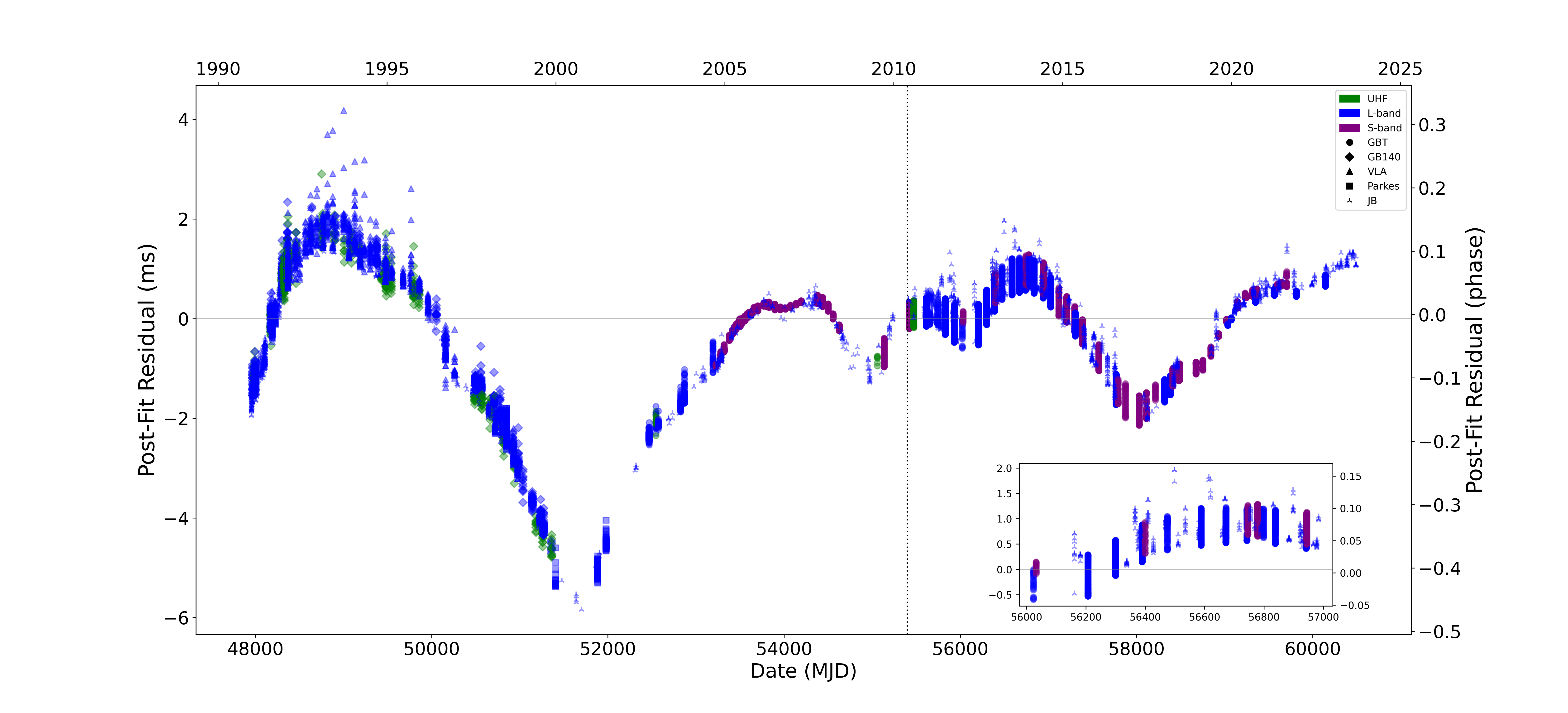}
    \caption{The long-term timing solution for Ter5A for $\Delta\mathrm{T}0=0$. The vertical dotted line represents where the processed TOAs switched from incoherent to coherent de-dispersion. For readability, only every tenth TOA is plotted. Note that the long-term behavior is smooth and phase-connected, while observations further from the center, which have less precise T0 values, show larger and larger systematics from errors in the predicted orbital phase. The inset is a zoom-in that clearly shows the spread of TOAs over individual observations and short-term variations.}
    \label{fig:simple}
\end{figure*}

\begin{figure*}[h!]
    \centering
    \includegraphics[width=\textwidth]{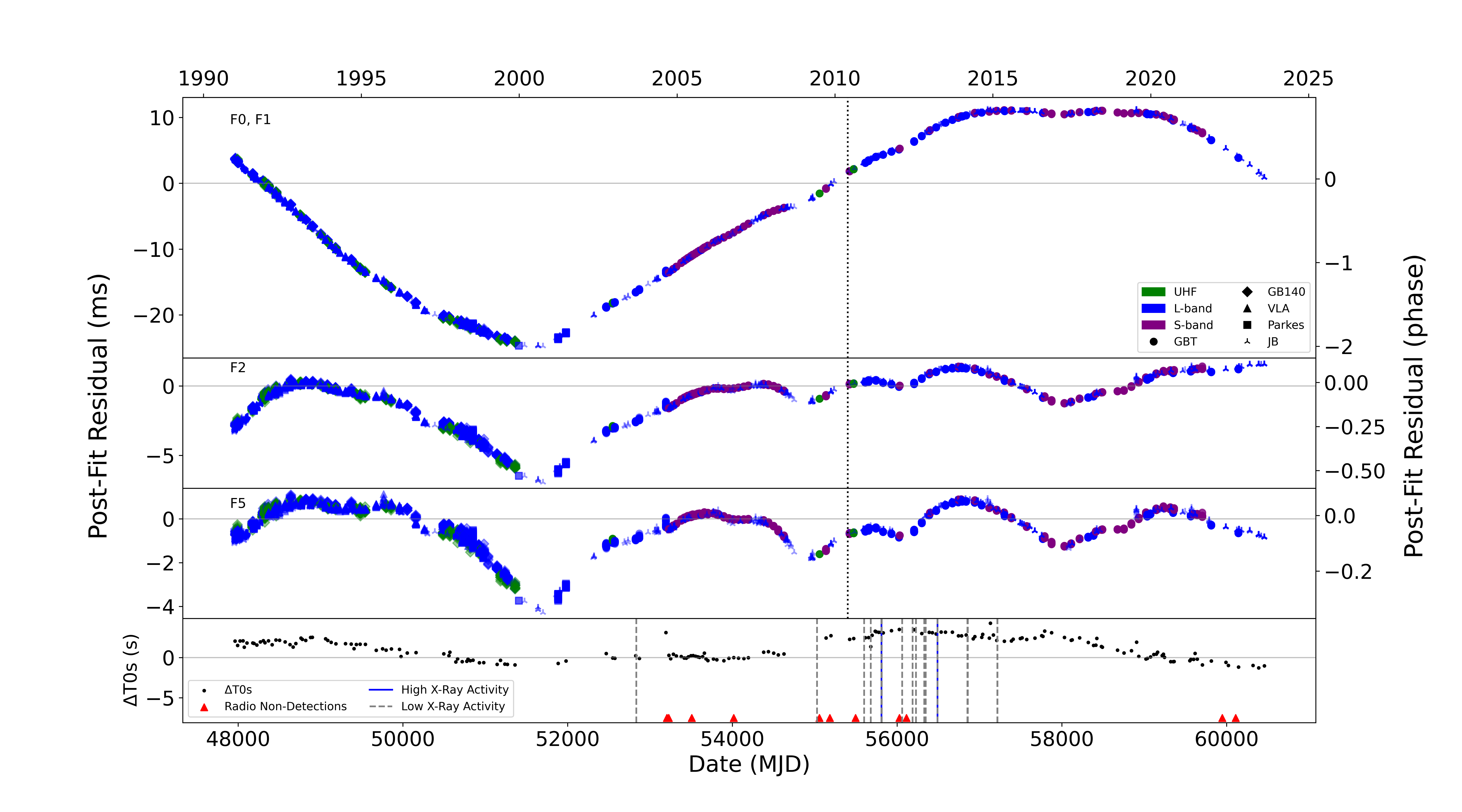}
    \caption{The top three panels show the timing residuals of three different Ter5A spin frequency derivative fits (i.e. F0$-$F1, F0$-$F2, and F0$-$F5, as indicated in the upper left hand corner of each panel). The bottommost panel shows orbital variations $\Delta$T0 with $\dot{P}_\mathrm{orb}$ removed (see sections \ref{T0var} and \ref{PBDOT} for more details) as well as all known X-ray detections (see section \ref{xray}) of Ter5A. The red triangles represent days in which the pulsar was observed but not detected in the radio in GBT data after MJD 53193 and in GB140 and VLA data. The vertical dotted line across the upper panels represents where the processed TOAs switched from incoherent to coherent de-dispersion. For readability, only every tenth TOA is plotted. Note the decreasing range covered by the y-axes in each case, showing nominally better fits. Strong systematic trends still remain in the residuals. There also remain strong DM and system \texttt{JUMP} effects, most evident in the data prior to the year 2000, which could be addressed in future work. It is apparent that X-ray activity, radio non-detections, orbital variations, and spin variations are not strongly correlated with each other.}
    \label{fig:FreqDerivPlots}
\end{figure*}

\subsection{Data Cleaning} \label{DataCleaning}

Before timing Ter5A, we removed the effects of the eclipse from observations because the time-variable ionized gas can cause unpredictable time delays in the measured pulses. To remove most of these effects, we cut all TOAs corresponding to an orbital phase between 0.0 and 0.5, where the pulsar was most likely to be eclipsed. We further inspected the data for additional eclipses and manually removed them. See Figure~\ref{fig:better_eclipse} for how eclipses appear in data before and after this cleaning; it is straightforward to visually identify affected TOAs and remove them. A summary of the phase-cleaned observations is presented in Table~\ref{tab:obssum}.

Figure~\ref{fig:better_eclipse} shows an example of what eclipses look like in the timing data and what the timing residuals look like after cleaning TOAs. As the pulsar moves behind a clump of ionized gas, this gas increases the effective dispersion measure (DM), causing pulse delays. The stochastic variability of the wind from the companion makes timing delays unpredictable. While every orbit has some eclipse around phase 0.25, the width and effect of this eclipse varies significantly between observations. Several observations show additional micro-eclipses or hours-long total disappearances. This is one reason that long-term timing of this system was previously prohibitively difficult.

\begin{table*}[t] 
    \centering 
    \caption{Summary of eclipse-cleaned observational data.} \label{tab:obssum} 
    \begin{tabular}{|c|c|c|c|D{.}{.}{-1}|c|}
    \hline Observatory & Num. Observations & Num. TOAs & Frequency (MHz) & \multicolumn{1}{c|}{Median Error ($\mu$s)} & Dates (MJD) \\ 
    \hline 
        GBT & 52 & 6279 & 820$-$2000 & 4.43 & 52466$-$55137 \\ 
        GBT* & 61 & 6477 & 820$-$2165 & 4.31 & 55422$-$60144 \\ 
        Green Bank 140\,ft & 91 & 1049 & 670$-$1600 & 23.7 & 47966$-$15363 \\ 
        Parkes & 4 & 619 & 1316$-$1454 & 24.3 & 50800$-$52000 \\ 
        VLA & 36 & 661 & 1667 & 11.1 & 48190$-$50975 \\ 
        Jodrell Bank & 219 & 1219 & 1396$-$1640 & 31.4 & 47952$-$60491 \\ 
        Jodrell Bank* & 68 & 520 & 1414$-$1725 & 15.6 & 55679$-$60321 \\ 
    \hline 
    \end{tabular} 
    \text{*Indicates data uses coherent de-dispersion. All other data are incoherently de-dispersed.} \\ 
   GBT data were coherently de-dispersed at a DM of 238\,pc\,cm$^{-3}$. JB data were coherently de-dispersed at a DM of just over 242.23\,pc\,cm$^{-3}$.
\end{table*}

\begin{figure}
    \centering
    \includegraphics[width=0.5\textwidth]{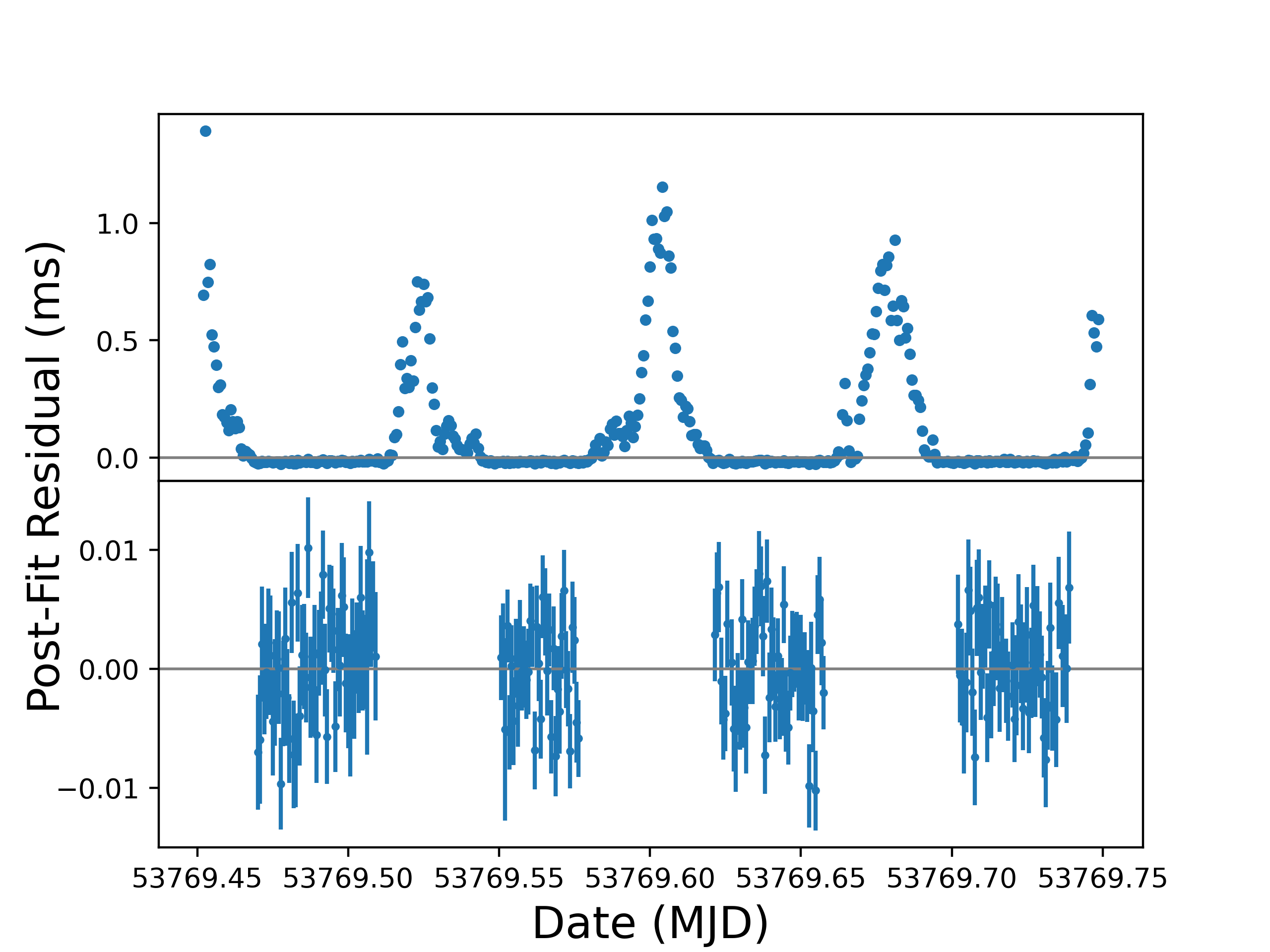}
    \caption{Timing residuals from a representative day (MJD 53769) before (top) and after (bottom) removing the effects of the eclipses. These data span four orbits. Note the vertical scales differ in the panels -- cleaning the eclipse reduces the residuals by two orders of magnitude. The eclipses are not perfectly symmetrical or identical from orbit to orbit. This reflects the rapidly-changing gas environment.}
    \label{fig:better_eclipse}
\end{figure}

\section{Timing Methodology} \label{timing}

Timing this system required new methodologies. The first new method, which we call the ``piecewise" method, is detailed in Section~\ref{piecewise}. While piecewise timing in general is not novel, and has been used to account for DM variations and pulsar glitches, we used a new piecewise orbital model for this timing \citep{O'Neill2025}. The second new method, which we call the ``isolation" method, is detailed in Section~\ref{iso}. While functions, in {\tt PRESTO}\footnote{\url{https://github.com/scottransom/presto}} \citep{ransom2001,presto} for instance, allowing the removal of the Roemer delay from barycentered events such as x-ray or gamma-ray photons have been used for short duration pulsation searches of specific systems, ``isolating" a long-duration topocentic timing dataset of a binary pulsar with radio TOAs is a novel and more complicated process, and it circumvents portions of the traditional radio pulsar timing software models.

The main reason that redback systems are difficult to time over long baselines is because the variable gas interactions and complicated companion effects cause random orbital period variations. These effects mean that timing solutions do not maintain phase coherence for more than a few years, or even a few months, beyond the end of any given timing solution. Phase-connection, which means successfully tracking every rotation of the pulsar, is the first step in generating a timing solution.

We measure orbital variations with the parameter ``T0." T0 is the precise time of the periastron of the orbit. For an effectively circular system such as this one, where eccentricity is set at zero, T0 is defined as the time of the ascending node, which is the instant when the pulsar crosses the plane of the sky moving away from the observer. For well-behaved binary systems, this value evolves uniformly over time according to Eqn~\ref{eqn:T0} (see section \ref{T0var}), but for redback systems, interaction with the companion drives orbital wander that results in unpredictable time-variable T0s. As a result, standard pulsar timing cannot easily account for this orbital behavior, which limits the duration that redback pulsars can maintain phase coherence. Most earlier spider pulsar timing used the so-called BTX timing model, which uses many terms in a Taylor expansion of the orbital frequency to account for this orbital wander. Such expansions can be highly numerically unstable and provide extremely poor extrapolations of the orbital phase. This exacerbates redback timing difficulties. In this paper we outline two different and effectively new methods (see sections \ref{piecewise} and \ref{iso}) by which we successfully timed Ter5A.

\subsection{Generating TOAs} \label{GenTOAs}

We searched each observation of the GBT and Parkes data for pulsations using the \texttt{spider\_twister}\footnote{\url{https://github.com/alex88ridolfi/SPIDER_TWISTER}} package, which searches over orbital phase by performing trial folds to determine a most probable T0. With an adequate local orbit, we then re-folded the observations with one minute integrations to determine TOAs, which could then be used to determine a much more precise T0 via timing software (e.g.~\texttt{Tempo} or \texttt{PINT}). For the GBT data, the folding and TOA determination were performed with the \texttt{prepfold} and \texttt{get\_TOAs.py} commands, respectively, from \texttt{PRESTO}. 

While this produced good starting parameters for individual days, we needed to use a model that accommodated time-varying T0s to determine a long-term timing solution. We independently used two different techniques to determine the same long-term timing solution.

\subsection{Piecewise Method} \label{piecewise}

Using \texttt{PINT}, we fit long-term timing parameters through a bootstrapping process, starting with initial parameters from published solutions \citep[spin frequency, frequency derivative, position, DM, orbital period, and semimajor axis; e.g.~][]{Nice1990,Nice2000Proceedings,Urquhart2020}, and phase-connecting the pulsar in 2$-$3 year segments.

We followed a four-step phase-connection process. First, we determined orbital phases (i.e.~the best T0) for each observation (see section \ref{GenTOAs}). Second, we phase-connected overlapping two-year chunks of data, fitting for frequency, first and second frequency derivatives, and position. Third, we verified that the number of pulse rotations between each TOA in the overlapping sections were identical for each phase-connected section. Fourth, we joined all the sections together using the rotational counts between all the TOAs.

This process produced a phase-connected timing solution. We next held the T0 values fixed while fitting for long-term parameters: orbital period, spin frequency, spin frequency derivatives one through five, DM, and DM derivative. \texttt{PINT} struggled with fitting the 10000+
TOAs with the \texttt{binary\_piecewise} model, given the large number of free parameters, which is one of the reasons we pursued an additional ``isolation" fitting technique (see section \ref{iso}). To accommodate non-GBT data, we manually fit for the \texttt{JUMPs} between datasets. We did this by fixing all long-term parameters and then isolating portions of the data that had two different instruments and fitting for JUMPs in these isolated portions in a pair-wise round-robin fashion until instrumental offsets were smaller than offsets due to T0 measurement errors and much smaller than overall timing noise. 

In order to nail down the short-term orbital variations within the long-term solution, we assumed that the orbital period was constant over roughly two to four-week periods (see Figure~\ref{fig:simple} and the top panel of Figure~\ref{fig:GPR}, which assume constant orbital period: the residuals are flat on timescales of a year or less and deviations from the constant orbital period model are only apparent on years-long timescales), and used multiple observations within such periods to constrain the piecewise constant T0 for those data. This was more important for the data before the year 2000, which had fewer and less precise TOAs.

The orbital phase residuals between the measured and predicted T0s, assuming a constant orbital period (see sections \ref{T0var} and \ref{PBDOT}), showed a stochastic-like pattern with a general quadratic trend. We used a least-squares fit to determine a best-fit orbital period derivative from the T0s, and also performed a Gaussian Process Regression which allowed us to predict and refine T0 measurements that were poorly determined in the first iteration.

The result is a phase-connected timing solution with relatively well-behaved TOAs in individual observations and relatively smooth long-term behavior, as well as a time series of orbital phase wander as determined by the measured values of T0 (Figures \ref{fig:FreqDerivPlots} and \ref{fig:GPR}).

\begin{figure*}
    \centering
    \includegraphics[width=\textwidth]{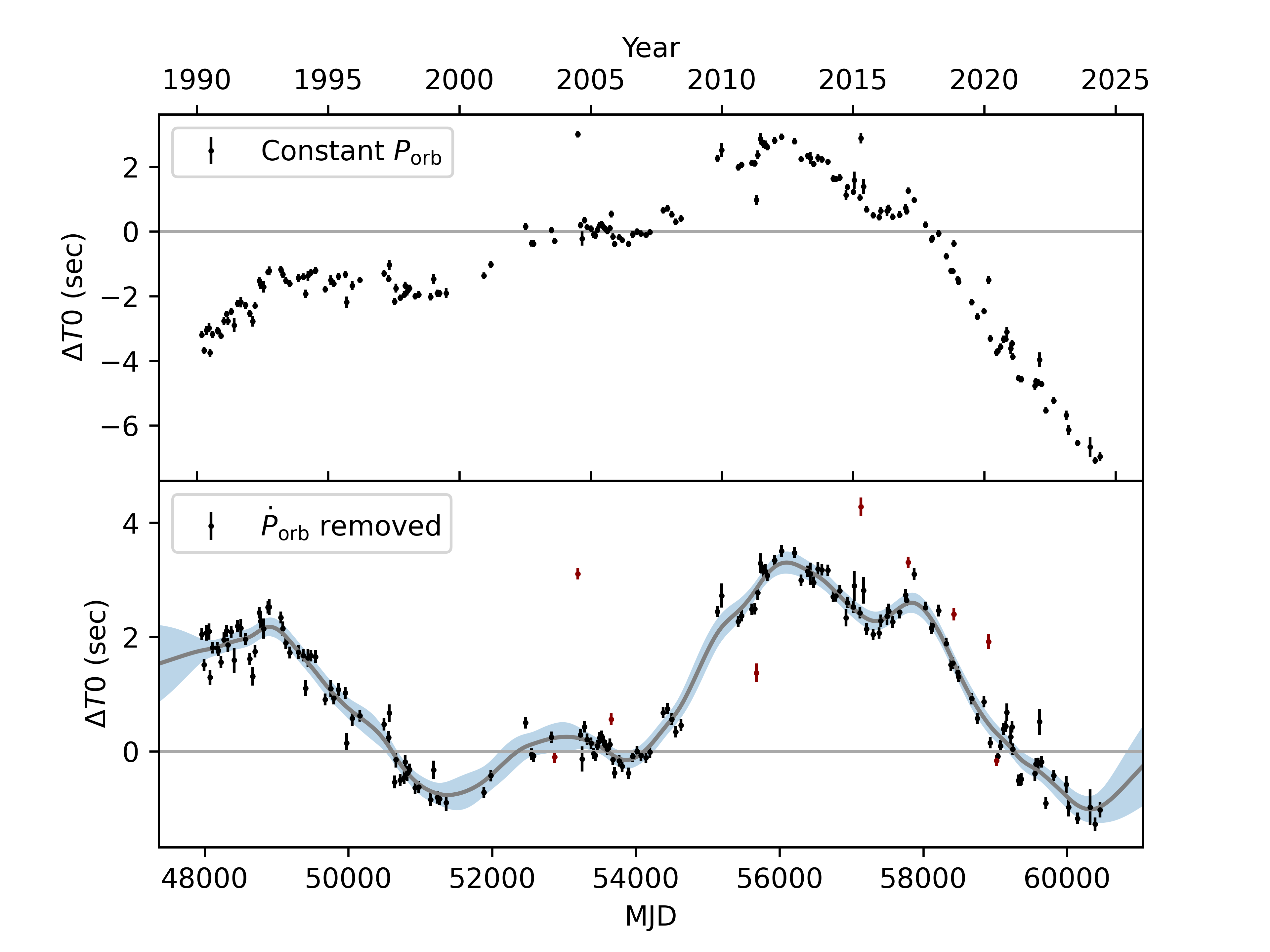}
    \caption{Time of ascending node deviations ($\Delta$T0) over time for Ter5A assuming a constant orbital period (top) and with the best-fit $\dot P_\mathrm{orb}$ removed (bottom).  The bottom panel shows a Gaussian process regression to the measured $\Delta$T0 values. Red data points are days in which the eclipse effects are covariant with T0 and prohibit accurate measurement, and are excluded from the GPR fit.}
    \label{fig:GPR}
\end{figure*}

\subsection{Isolation Technique} \label{iso}

Besides this piecewise technique, and as part of a related project on long-term timing of other globular cluster redbacks \citep{Corcoran2024}, we also phase-connected the system without breaking it into chunks by removing the effects of the binary and effectively ``isolating'' the pulsar. We used \texttt{PINT} and the accurately known T0 values measured over short timescales (see Table~\ref{tab:T0s_all}), and the long-term $P_\mathrm{orb}$, and $\dot{P}_\mathrm{orb}$ values to subtract off the Roemer delays from the binary orbit from the TOAs for each observation, resulting in a much simpler dataset: an ``isolated'' pulsar with far fewer parameters to fit.

This isolation technique effectively de-couples the orbital variations from the long-term spin behavior of the pulsar. This decoupling is possible because the Fourier components of the orbital and spin periods are separated by many orders of magnitude. For an explanation of why timing effects with radically different periods do not significantly contaminate each other, see Section 4.2 of \cite{Bochenek2015}. Since the arrival times are topocentric (i.e.~in the reference frame of the observatories), but the binary Roemer delays need to be computed in the barycentric frame, we had to perform a first-order correction to the predicted Roemer delays to compensate for the difference in those frames. We tested this technique with simulated binary TOAs and were able to to remove all orbital effects to better than 100\,ns. It is likely to be useful in other long-term redback timing efforts, and can be implemented in \texttt{PINT} with just a few lines of code.

Ultimately we used the short-term timing position measured with the GBT Spigot data (see Figure~\ref{fig:positions} and Table~\ref{tab:positions}) rather than a long-term timing-fitted position in our final model, and \textit{Gaia}-measured proper motion values \citep{ClusterPM} for the cluster rather than the timing-fitted proper motion of Ter5A (see section~\ref{pospm}). One of the reasons for this was that the VLA-determined astrometric positions are not precise enough in RA for our purposes. For further discussion of position complications, see Section~\ref{pospm}. The long-term timing parameters are presented in Table~\ref{timingtab}.

\section{Results}

\subsection{Simplest Long-Term Model} \label{simplelongterm}

If we do not accommodate for changing T0, and just use pulse numbers to fit for spin frequency and first and second spin frequency derivatives, we obtain a phase-connected timing solution with a smooth overall trend. However, within any individual observation the pulse phase varies by 5$-$10\%. This solution is shown in Figure~\ref{fig:simple} and in Table~\ref{timingtab}.

\subsection{Spin Frequency Derivatives} \label{sfd}

We fit the data through five spin frequency derivatives to show the extent of the spin noise in the data. Figure~\ref{fig:FreqDerivPlots} shows the evolution of fit with successive derivatives. All plots shown have been fit for DM, DM derivative, frequency, and the number of frequency derivatives indicated. We denote spin frequency as F0, spin frequency derivative as F1, second spin frequency derivative as F2, and so on.

Ter5A is a messy system and over a 34-year timespan the DM changes significantly, both on long time scales and, on some days, on minute-to-hour timescales. In principle, a separate DM could be used for each observation. However, this is impractical because of the number of free parameters involved, the occurrences of DM variations within individual observations, and a lack of wide band data suitable for DM measurements in many observations. As a result, DM is not perfectly modeled in the present analysis. However, compared to the timing noise systematics, the variations due to the DM are small (i.e.~few percent of a pulse period for the F1-based solution and up to $\sim$10\% of a rotation for the F5-based solution). The only time we see large excess DM delays ($\gtrsim 1$\,pc\,cm$^{-3}$) on short time scales is during an eclipse. There is very little evidence of days-to-weeks long variations of a similar size, except for on a very small number of observations in which the excess DM is clear in visual inspection of the data and results in inaccurate T0 measurements. These days are marked in red in Figure~\ref{fig:GPR}. On all other days, removing the eclipse and measuring a local T0 results in flattened residuals. In Figure~\ref{fig:simple}, only a handful of TOAs out of about twenty thousand are significantly DM-delayed, and these TOAs were not part of the phase connection process. Therefore, DM variations are not a significant source of error in our results.

There is an apparent abrupt timing feature around the year 2000 (i.e.~MJD 51700). This is noted in \cite{Nice2000Proceedings} and corroborated by our observations. While it is unfortunate that this feature occurs during the least-well sampled period of our data, the unambiguous phase connection of the Parkes and Jodrell Bank observations during that time period suggests this feature is real and corresponds to some physical change in the system. We note that whatever this physical change is, it appears strongly only once the second frequency derivative has been subtracted. We also note that several other timing features of smaller amplitude appear later in the data (e.g.~around MJDs 55000, 56000, and 58000).

We measure a spin frequency derivative (i.e.~F1) of $\sim$1.35$\times$10$^{-16}$\,Hz\,s$^{-1}$, depending on the exact spin model used. This positive value is consistent with other measured values \citep[e.g.~][]{Discovery,Nice1992,Prager2017}. The intrinsic spin-down rate of the pulsar must be negative. The positive measured first spin derivative value is likely due to the globular cluster accelerating the pulsar toward the observer from the far side of the cluster. There are additional mechanisms that contribute to apparent positive F1 values, and they usually affect the orbit and spin equally. We discuss this in the next two sections.

Whether the spin and orbital variations (see Figure~\ref{fig:FreqDerivPlots}) are correlated is a tricky matter. Red noise dominates in both the spin residuals (topmost panel) and orbital residuals (bottommost panel). By definition red noise dominates at the lowest frequencies, and since these residuals have subtracted out both the linear and quadratic terms (F0 and F1, and $P_\mathrm{orb}$ and $\dot{P}_\mathrm{orb}$, respectively), red noise dominates the remaining cubic. The dominating cubic has the same sign in both the spin and orbital residuals. This creates a very rough overall correlation that is inconclusive - any two random cubic functions have a 50\% chance of having the same sign, and therefore a potentially spurious correlation, and a 50\% chance of having the opposite sign, and a potentially spurious anticorrelation. Definitive evidence of correlation can be sought only in shorter-term regions of the data set. We note that on these smaller timescales, where there are dramatic turnarounds in the spin residuals, we see no definitive correlated changes in the orbital residuals. For a few specific examples, at the time of the dominant change in the spin behavior of the pulsar just after the year 2000, the orbital variations might show a slight positive correlation. However, at the next two biggest spin residual changes, near 2009 and 2016, there seems to be little to no change in the orbital variations or a negative correlation, respectively.

X-ray activity does not appear to correlate with either of these variations. Similarly, the days on which the pulsar was undetected in radio observations also do not appear to correlate with those variations. 

The lack of radio non-detections from MJD 56000 to MJD 60000 is a little surprising. It is possible that the amount of intra-binary activity fluctuates on wide varieties of timescales, and so there may be long periods where the pulsar is a well-behaved eclipsing system, and then other periods where its visibility is much more variable. 

The total fractional sensitivity of the pulsar's spin (median spin phase error divided by the total pulse phase covered by the dataset, T$\times$F0) is $4.4\times10^{-15}$ while the total fractional sensitivity of the pulsar's orbit (median T0 phase error divided by the total orbital phase covered by the dataset, T$\times P_\mathrm{orb}^{-1}$) is $9.3\times10^{-11}$,  so the spin residuals are substantially more sensitive to torques and other effects on the system than the T0 residuals.

We examine the spectrum of the spin residuals by taking a Fourier decomposition of the F0$-$F1 solution (Figure \ref{fig:FreqDerivPlots}, top panel) using a single ``median residual'' TOA from each observation. We calculated the powers of the components using the \texttt{get\_wavex\_amps()} and \texttt{get\_wavex\_freqs()} functions of \texttt{pint.utils} and determined a best-fit line using \texttt{scipy.curve\_fit()}. This gives a steep power-law-like spectrum with a spectral index of $-5.54(7)$, as shown in Figure \ref{fig:spinwaves}. The error represents the span between the maximum and minimum slope values estimated by multi-variate gaussian sampling using the fit covariance matrix. The first two harmonics are the most significant and dominate the measured spectral index; however, they are also potentially underestimated. The first bin is partially covariant with F0 and F1, which had already been subtracted from the residuals. The second bin is also partially covariant with these parameters, although to a lesser extent.

\renewcommand{\arraystretch}{0.9}
\begin{longtable}{cccccccccccc}
    \label{tab:T0s_all}\\
    \caption{Our final measurements for $T_{0}$ for Ter5A.}\\

    \toprule

    \multicolumn{1}{c}{{Group \#}} &
    \multicolumn{1}{c}{\scriptsize{$T_0$ [MJD]}} &
    \multicolumn{1}{c}{{Group \#}} &
    \multicolumn{1}{c}{\scriptsize{$T_0$ [MJD]}} &
    \multicolumn{1}{c}{{Group \#}} &
    \multicolumn{1}{c}{\scriptsize{$T_0$ [MJD]}} \\

    \midrule

    1  & 47967.5211462(25)  & 49  & 51142.691189(53)   & 97  & 55653.468991(86)   \\
    2  & 47992.8625959(33)  & 50  & 51184.7504350(17)  & 98  & 55743.336581410(36)\\
    3  & 48182.8856352(22)  & 51  & 51236.9462491(14)  & 99  & 55829.951378509(43)\\
    4  & 48225.85262672(85) & 52  & 51265.918711(13)   & 100 & 55931.61975729(14) \\
    5  & 48268.6683358(10)  & 53  & 51383.2458160(22)  & 101 & 56027.76596080(11) \\
    6  & 48301.0448706(16)  & 54  & 51884.0231145(21)  & 102 & 56206.895965627(46)\\
    7  & 48324.9490412(36)  & 55  & 51979.94239029(41) & 103 & 56299.638095659(43)\\
    8  & 48366.4031143(44)  & 56  & 52468.01112700(41) & 104 & 56389.35438867(24) \\
    9  & 48457.5566862(48)  & 57  & 52546.38049423(66) & 105 & 56399.33967526(20) \\
    10 & 48501.960954(85)   & 58  & 52573.91567960(43) & 106 & 56474.229326535(27)\\
    11 & 48568.832117(25)   & 59  & 52826.19547329(49) & 107 & 56587.849792656(53)\\
    12 & 48633.0556662(72)  & 60  & 52869.994569(90)   & 108 & 56671.590041286(69)\\
    13 & 48698.413910(15)   & 61  & 53193.3060992(25)  & 109 & 56743.907724038(85)\\
    14 & 48759.2333977(51)  & 62  & 53228.02763380(11) & 110 & 56782.714179(50)   \\
    15 & 48821.1875594(31)  & 63  & 53281.887669276(59)& 111 & 56798.145987630(65)\\
    16 & 48877.9221511(45)  & 64  & 53319.78637056(16) & 112 & 56838.162781910(43)\\
    17 & 48901.9019577(25)  & 65  & 53378.639047620(88)& 113 & 56942.857002491(48)\\
    18 & 49004.705062(14)   & 66  & 53414.495304212(86)& 114 & 57026.67289664(98) \\
    19 & 49004.7050624(89)  & 67  & 53438.399476352(89)& 115 & 57119.339385812(59)\\
    20 & 49061.4396263(21)  & 68  & 53466.31289481(14) & 116 & 57209.131320225(42)\\
    21 & 49086.8567026(15)  & 69  & 53503.30384691(29) & 117 & 57302.856855201(45)\\
    22 & 49128.3107882(70)  & 70  & 53522.820545107(97)& 118 & 57387.656147661(46)\\
    23 & 49184.591495(70)   & 71  & 53553.15463624(13) & 119 & 57573.140418710(33)\\
    24 & 49236.938586(23)   & 72  & 53579.10125297(22) & 120 & 57769.5933781(24)  \\
    25 & 49303.8097748(13)  & 73  & 53600.96297964(13) & 121 & 57791.6064045(17)  \\
    26 & 49372.1182084(61)  & 74  & 53631.0701342(68)  & 122 & 57875.271003766(80)\\
    27 & 49434.9044942(17)  & 75  & 53659.059202373(62)& 123 & 58030.87505377(30) \\
    28 & 49480.7460397(19)  & 76  & 53678.878476315(45)& 124 & 58113.6318991(62)  \\
    29 & 49544.6670191(14)  & 77  & 53702.782646032(46)& 125 & 58215.30028134(11) \\
    30 & 49676.2912430(14)  & 78  & 53769.57816795(35) & 126 & 58320.070144442(90)\\
    31 & 49781.5906512(15)  & 79  & 53811.4861145(21)  & 127 & 58384.898861061(76)\\
    32 & 49858.7496867(23)  & 80  & 53830.47328809(96) & 128 & 58412.812278032(51)\\
    33 & 49958.5269046(47)  & 81  & 53896.4366994(11)  & 129 & 58486.113361403(61)\\
    34 & 50052.78196093(90) & 82  & 53956.95359396(65) & 130 & 58676.13639872(59) \\
    35 & 50157.930063(43)   & 83  & 54015.42804086(67) & 131 & 58753.9006012(15)  \\
    36 & 50261.3383028(15)  & 84  & 54073.448609(35)   & 132 & 58844.675942632(60)\\
    37 & 50492.9667102(12)  & 85  & 54138.42861983(88) & 133 & 58933.408827089(39)\\
    38 & 50564.3009975(60)  & 86  & 54194.48239060(60) & 134 & 59029.101158831(54)\\
    39 & 50663.5486979(17)  & 87  & 54379.9666671(11)  & 135 & 59076.001752235(65)\\
    40 & 50714.458531(35)   & 88  & 54437.98723631(11) & 136 & 59146.80651910(22) \\
    41 & 50730.8737453(16)  & 89  & 54500.546569301(67)& 137 & 59240.00252734(85) \\
    42 & 50782.3130975(39)  & 90  & 54556.37339788(13) & 138 & 59317.388495254(21)\\
    43 & 50818.1693568(61)  & 91  & 54626.270407(31)   & 139 & 59356.724475713(72)\\
    44 & 50852.21011121(87) & 92  & 55058.966185250(56)& 140 & 59565.734689747(72)\\
    45 & 50925.889426(41)   & 93  & 55136.806040349(63)& 141 & 59599.69979484(13) \\
    46 & 50938.2197420(23)  & 94  & 55423.050929003(71)& 142 & 59644.4066483(12)  \\
    47 & 50979.7494554(32)  & 95  & 55472.901718495(35)& 143 & 59705.30176143(95) \\
    48 & 51035.9545259(15)  & 96  & 55614.43559669(62) & 144 & 59814.0808785(20)  \\

    \bottomrule
      
\end{longtable}
\renewcommand{\arraystretch}{1}

\subsection{Orbital Variations} \label{T0var}

In the absence of external effects, such as that of accelerations from the globular cluster, tidal effects in the bloated companion star, gravitational wave radiation from the compact orbit, and mass transfer or loss from the companion, we can expect T0 will evolve according to 
\begin{equation}
    \mathrm{T}0_n = \mathrm{T}0_{0} + n \times P_\mathrm{orb},
    \label{eqn:T0}
\end{equation}
where T0$_{0}$ is some reference measured T0, $n$ is an integer number of orbits that have elapsed between T0$_{0}$ and T0$_{n}$, and $P_{\rm orb}$ is the orbital period of the binary (1.82\,hours). We define variations from a constant $P_{\rm orb}$ model with the quantity $\Delta\mathrm{T}0$:
\begin{equation}
    \Delta\mathrm{T}0 \equiv \mathrm{T}0 - \mathrm{T}0_n,
\end{equation}
where T0 is the measured T0 and T0$_n$ is the predicted T0 for a given observation.
When we compare our measured T0 values with those predicted by Eqn~\ref{eqn:T0} using a precise estimate for $P_\mathrm{orb}$ (see Table \ref{timingtab}), we find the system deviates from a constant-$P_{\mathrm{orb}}$ model by up to $\Delta\mathrm{T}0 \sim 10$\,s, as shown in Figure~\ref{fig:GPR}. There is a quadratic trend corresponding to an orbital period derivative ($\dot P_\mathrm{orb}$). Adjusting for the effects of $\dot P_\mathrm{orb}$, as described in the next section, results in a random process varying across $\sim$5\,seconds, also shown in Figure~\ref{fig:GPR}. 

We also performed a Gaussian Process Regression on the measured T0 values (see Figure~\ref{fig:GPR}) to estimate how smooth the variations might be in time and to better estimate T0 values on days where it could not be measured well. On several days, micro-eclipses or significant ionized gas effects during the observation cause strong systematic offsets in the T0 measurements due to unmodeled DM changes. Our final values are presented in Table~\ref{tab:T0s_all}.

The spread in the daily residuals of Figure~\ref{fig:simple} due to an error in orbital phase can be explained by the deviation from the straightforward T0 evolution (Eqn~\ref{eqn:T0}) shown in the top panel of Figure~\ref{fig:GPR}. For a circular orbit with speed $v_{\mathrm{orb}}$, the magnitude of residual TOAs due to an error in orbital phase, $\Delta T0$, is $v_{\mathrm{orb}}\Delta T0$. For Ter5A, $v_{\mathrm{orb}}=2\pi(a_1\sin i)/P_{\mathrm{orb}}=1.14\times 10^{-4}~{\mathrm{s}}/{\mathrm{s}}$, where $a_1\sin i$ is the orbital semi-major axis projected onto the line of sight, and $v_{\mathrm{orb}}$ is also projected onto the line of sight. For a typical deviation of $\Delta T0=2~{\mathrm{s}}$, (Figure~\ref{fig:GPR}), this gives residuals of magnitude 0.23~ms, consistent with the spread in the residuals of any given observation (Figure~\ref{fig:simple}). Where $\Delta T0=4\mathrm{-}8~{\mathrm{s}}$, the observed residual spread in an observation is likewise 0.46-0.92\,ms.

We determined the Power Spectral Density (PSD) (see Figure~\ref{fig:PSD}) of these $\dot{P}_\mathrm{orb}$ subtracted orbital variations by using \texttt{scipy.curve\_fit()} to take a best fit line of the portion of the PSD between the first bin ($8.0 \times 10^{-5}$\,d$^{-1}$) and where the spectrum turns into white noise ($10^{-3}$\,d$^{-1}$). The first bin is partially covariant with $\dot P_\mathrm{orb}$ and therefore has a reduced power. This produces a power-law with a spectral index $\gamma$ of $-0.9(2)$.

Three other works have assessed the spectral index of T0 variations in redbacks. From an analysis of Fermi-LAT observations of three redback pulsars, \cite{Thongmeearkom2024} found orbital phase wander spectral indices constrained to $\gamma < -2.4$, $\gamma = -3.81^{+0.32}_{-0.48}$, and $\gamma < -5.4$. \cite{Corcoran2024} surveys five redbacks and measures spectral indices of $\gamma = -0.7(2)$, $-0.7(2)$, $-0.9(1)$, $-1.1(2)$, and $-1.7(2)$. \cite{Clarke2021}'s analysis concluded there were too few significant frequency bins for a meaningful measurement. Our measurement of $\gamma = -0.9(2)$ is shallower than the indices observed in \cite{Thongmeearkom2024} and consistent with those observed in \cite{Corcoran2024}. 

\begin{figure*}
    \centering
    \includegraphics[width=\textwidth]{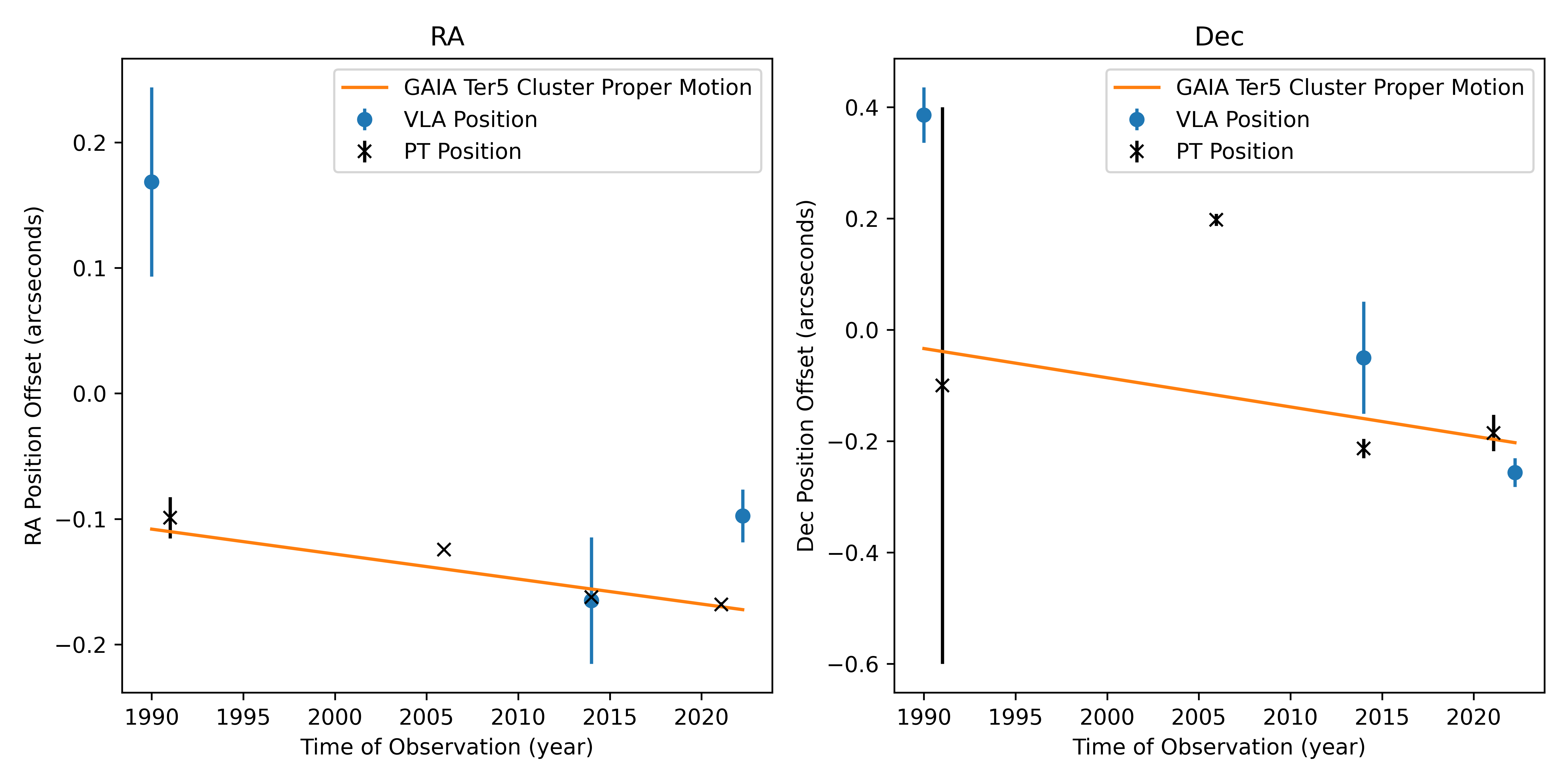}
    \caption{A comparison of pulsar timing and VLA-measured positions of Ter5A with the \textit{Gaia} proper motion of the cluster. The four timing positions are from \cite{Nice1992}, and three-year sections of SPIGOT, GUPPI, and VEGAS observations (Figure~\ref{fig:zoom}). The VLA positions are from \cite{FG2000}, \cite{Urquhart2020}, and Urquhart et al. (in prep), and their error bars are $1\sigma$. Note both right ascension and declination are in units of arcseconds.}
    \label{fig:positions}
\end{figure*}

The Applegate model \citep{Applegate} predicts that spider pulsar systems should show ``regular, but not strictly periodic'' modulations in orbital period due to cycles of magnetic activity and changing quadrupole moment of the companion star. We see no strong evidence for such periodicity on the scale of up to $\sim$20 years.

\subsection{Orbital Period Derivative} \label{PBDOT}

The orbital period change due to gravitational wave emission \citep{PetersMathews} can be expressed as a function of a system's orbital parameters (e.g., \citealt{TaylorWeisberg,FreireWex2024}):
\begin{equation}
\begin{split}
  \dot P_\mathrm{orb} = -\,\frac{192\pi}{5}
  \left(\frac{P_\mathrm{orb}}{2\pi}\right)^{-5/3}
  \left(1 + \frac{73}{24} e^2 + \frac{37}{96} e^4 \right) \\
  \times (1-e^2)^{-7/2}\,T_\odot^{5/3}\, m_1\, m_2\, M^{-1/3},
  \label{eq:pbdot} 
\end{split}
\end{equation}
where $e$ is the eccentricity, $P_\mathrm{orb}$ is the orbital period, $M$ is the total mass of the system, $m_1$ and $m_2$ are the component masses of the binary, and $T_\odot\equiv GM_\odot/c^3 = 4.925490947\,\mu$s. For Ter5A, the eccentricity is zero, simplifying Eqn~\ref{eq:pbdot} to
\begin{equation}
    \dot P_\mathrm{orb} = -\,\frac{192\pi}{5}
  \left(\frac{P_\mathrm{orb}}{2\pi}\right)^{-5/3}
  \,T_\odot^{5/3}\, m_1\, m_2\, M^{-1/3}.
  \label{eq:adj}
\end{equation}

The observed pulsar spin-down to spin period ratio $\left(\dot{P}/P\right)_\mathrm{spin,obs}$ can be converted to an acceleration:

\begin{equation}
    A_\mathrm{spin,obs} = c \left(\dot{P}/P\right)_\mathrm{spin,obs}.
\end{equation}

$A_\mathrm{obs}$ is the sum of several accelerations and acceleration-like terms that we will, for simplicity, refer to as accelerations:

\begin{equation} \label{eq:accelspin}
    A_\mathrm{spin,obs} =  A_\mathrm{int} + A_\mathrm{Shk} + A_\mathrm{GC} + A_\mathrm{gal}, 
\end{equation}
where $A_\mathrm{int} = c \left(\dot{P}/P\right)_\mathrm{int}$ is the intrinsic spin-down rate of the pulsar, $A_\mathrm{GC}$ is the contribution due to acceleration of the pulsar by the globular cluster, and $A_\mathrm{gal}$ is the contribution due to the acceleration of the pulsar by the Galactic gravitational potential. $A_\mathrm{Shk}$ is the apparent acceleration due to the Shklovskii effect which is given by 
\begin{equation}
    {A_\mathrm{Shk}} = \mu^2{d}
\end{equation}
where $\mu$ is the proper motion of the pulsar and $d$ is the distance to the pulsar. Of these, $A_\mathrm{GC}$ is the dominant term since the pulsar is observed to have a negative period derivative. 

Orbital period accelerations follow a similar form to Eqn. \ref{eq:accelspin}:
\begin{equation} \label{eq:accelorb}
    A_\mathrm{orb,obs} =  A_\mathrm{Shk} + A_\mathrm{GC} + A_\mathrm{gal} + A_\mathrm{GR} + A_\mathrm{RB}.
\end{equation}

The observed orbital acceleration, denoted by $A_\mathrm{orb,obs}$, is given by $c \times \left(\dot P_\mathrm{orb}/P_\mathrm{orb}\right)_\mathrm{obs}$. The term $A_\mathrm{GR}$ is an acceleration-like term representing the quantity $c \times \left(\dot{P}_\mathrm{orb}/P_\mathrm{orb}\right)_\mathrm{GR}$, which is the GR-predicted contraction due to gravitational wave emission. The apparent acceleration due to variations in the companion in the redback system is contained in $A_\mathrm{RB}$, which is not directly measurable, but can be constrained by the span of reasonable $A_\mathrm{GR}$ values, in combination with the estimates of the other acceleration parameters.

\begin{table}[h]
    \centering
    \footnotesize
    \begin{tabular}{p{2cm} c @{\extracolsep{5pt}}d{7} @{\extracolsep{5pt}}d{4} @{\extracolsep{5pt}}c}
         \hline
         Paper &  Method & \multicolumn{1}{c}{RA (s)} & \multicolumn{1}{c}{Dec (\arcsec)} & Epoch (MJD) \\
         \hline
         \citet{Nice1992} & Timing & 02.2534(11) & 37.7(5) & 48270 \\
         This work & Timing & 02.25170(3) & 37.40(2) & 53711 \\
         This work & Timing & 02.24918(5) & 37.81(2) & 56663 \\
         This work & Timing & 02.249(2)   & 37.7(4)  & 59374 \\
         \citet{FG2000} & VLA & 02.2685(50) & 37.23(5) & 47892* \\
         \citet{Urquhart2020} & VLA & 02.249(50)  & 37.65(10) & 56663* \\
         Urquhart et al. (in prep) & VLA & 02.2534(14) & 37.85(2) & 59669 \\
         \hline
    \end{tabular}
    \caption{J2000 measured positions of Ter5A represented in Figure~\ref{fig:positions}. Listed positions are the RA value in seconds of arc offset from 17h\,48m and the declination value in arcseconds offset from $-$24${\degree}$\,46\arcmin. Note the timing errors are statistical and reported VLA errors are $1\sigma$. The reported uncertainty on the position measured by \citet{FG2000} does not include  the systematic uncertainty from the position of the phase calibrator used or from phase interpolation uncertainties, and hence is potentially underestimated. \\ *Approximate value}
    \label{tab:positions}
\end{table}

\let\cleardoublepage\clearpage

\begin{deluxetable*}{lcc}
\tabletypesize{\footnotesize}
\tablecaption{Timing Parameters\label{timingtab}}
\tablewidth{0pt}
\tablehead{\colhead{Parameter} & \colhead{Simple F2} & \colhead{F5}}
\startdata
Pulsar Name \dotfill & PSR J1748$-$2446A & $-$ \\
\cutinhead{Timing Parameters}
Right Ascension (RA, J2000) \dotfill & $17^{\rm h}\;48^{\rm m}\;02\fs24918(4)$ & $-$ \\
Declination     (DEC, J2000) \dotfill & $-24\degree\;46\amin\;37\farcs81(2)$ & $-$ \\
Proper Motion\tablenotemark{a} in RA (mas\,yr$^{-1}$) \dotfill & $-$2.0 & $-$ \\
Proper Motion\tablenotemark{a} in DEC (mas\,yr$^{-1}$) \dotfill & $-$5.2 & $-$ \\
Position Epoch (POSEPOCH, MJD) \dotfill & 56663.0 & $-$ \\
Pulsar Spin Period (ms) \dotfill & $11.5631483824071(6)$ & $-$ \\
Pulsar Spin Frequency (Hz) \dotfill & $86.481636914861(5)$ & $86.48163691333(7)$ \\
Spin Frequency Derivative (Hz\, s$^{-1}$) \dotfill & $1.31969(9) \times 10^{-16}$ & $1.446(8) \times 10^{-16}$ \\
Frequency 2nd Derivative (Hz\, s$^{-2}$) \dotfill & $1.433(1) \times 10^{-25}$ & $2.29(6) \times 10^{-25}$ \\
Frequency 3rd Derivative (Hz\, s$^{-3}$) \dotfill & $-$ & $-1.4(1) \times 10^{-33}$ \\
Frequency 4th Derivative (Hz\, s$^{-4}$) \dotfill & $-$ & $-4.2(4) \times 10^{-42}$ \\
Frequency 5th Derivative (Hz\, s$^{-5}$) \dotfill & $-$ & $9.3(8) \times 10^{-50}$ \\
Reference Epoch (PEPOCH, MJD) \dotfill & 54221.719971 & $-$ \\
Dispersion Measure (DM, pc\,cm$^{-3}$) \dotfill & $242.36(2)$ & $-$ \\
DM Derivative (pc\,cm$^{-3}$\,yr$^{-1}$) \dotfill & $0.014(2)$ & $-$ \\
Span of Timing Data (MJD) \dotfill & 47952$-$60491 & $-$ \\
\cutinhead{Orbital Parameters}
Orbital Period (days) \dotfill & $0.07564611426(5)$ & $-$ \\
Orbital Period Derivative \dotfill & $-2.5(3) \times 10^{-13}$ & $-$ \\
Projected Semi-Major Axis (lt-s) \dotfill & $0.119624(2)$ & $-$ \\
Ref.~Epoch of Periastron ($T0$, MJD) \dotfill & $54015.8062583(1)$ & $-$ \\
Orbital Eccentricity \dotfill & 0 & $-$ \\
Longitude of Periastron, ($\omega$, deg) \dotfill & 0 & $-$ \\
\cutinhead{Derived Parameters}
Mass Function (\msun) \dotfill & $0.00032119(2)$ & $-$ \\
Min Companion Mass (\msun) \dotfill & $\geq$\,0.089 & $-$ \\
\enddata
\tablenotetext{a}{The proper motion was fixed at the GAIA proper motion of the cluster.}
\tablecomments{Numbers in parentheses represent 1-$\sigma$ uncertainties in the last digit
as determined by {\tt TEMPO}, {\tt PINT}, or via standard error propagation,
although many of the timing parameters are dominated by systematic errors.
The timing solutions used the DE440 Solar System Ephemeris and times are all in 
Barycentric Dynamical Time (TDB), referenced to TT(BIPM2021). Minimum companion mass 
was calculated assuming a pulsar mass of 1.4\,\msun. The DM reference epoch is the same as the PEPOCH for both solutions.}
\end{deluxetable*}

By combining Eqns \ref{eq:accelspin} and \ref{eq:accelorb}, we can eliminate the dependency on $A_\mathrm{gal}$, $A_\mathrm{GC}$, and $A_\mathrm{Shk}$ (for a breakdown of the predicted contributions of these terms, see Table \ref{tab:acc}):
\begin{equation} \label{eq:obsint}
    A_\mathrm{spin,obs} - A_\mathrm{orb,obs} = A_\mathrm{int} - A_\mathrm{GR} - A_\mathrm{RB}.
\end{equation}

The quantity $A_\mathrm{spin,obs} = c\left(\dot{P}/P\right)_\mathrm{spin,obs}$ has a $\sim$10\% fractional systematic error due to changes in the measurement of F1 as higher order frequency derivatives are fit (see Table~\ref{timingtab}). From fitting linear (i.e.~$P_\mathrm{orb}$) and quadratic (i.e.~$\dot P_\mathrm{orb}$) terms to the raw T0 residuals (Figure~\ref{fig:GPR}, top) we measure $\dot{P}_\mathrm{orb} = -2.5(3) \times 10^{-13}$.

Assuming a reasonable age for the pulsar of $5.0 \times 10^{9}$\,yr, within a factor of ten, and a surface magnetic field strength $B$ of $8 \times 10^{8}$\,G based on the mean of the 16 closest-period pulsars in the ATNF catalogue\footnote{\url{https://www.atnf.csiro.au/research/pulsar/psrcat/}}, excluding globular cluster pulsars, we can predict $\left(\dot{P}/P\right)_\mathrm{int}$. Varying $B$ across a reasonable range of $6 \times 10^{8}$ to $10^{9}$\,G, we find $\left(\dot{P}/P\right)_\mathrm{int} = 4.7^{+2.6}_{-2.0} \times 10^{-18}$, resulting in an intrinsic acceleration of $A_\mathrm{int} = 14.0^{+7.8}_{-6.0} \times 10^{-10}$\,m\,s$^{-2}$. 

To calculate $A_\mathrm{GR}$, we consider a 1.5\,\msun\ pulsar with a 0.09\,\msun\ companion star, which are reasonable estimates for the masses of the system components. MSPs are typically more massive than canonical pulsars, and the system is highly likely to have a large inclination due to the strong eclipses, meaning 0.09\,\msun\ is a likely companion mass. Assuming $A_\mathrm{RB} = 0$ yields the prediction shown in Table~\ref{tab:acc} for a predicted $\dot{P}_{\mathrm{orb}}$ of $-2.34^{+0.22}_{-0.17} \times 10^{-13}$, which is consistent with our measured value of $-2.5(3) \times 10^{-13}$. The error range on the predicted value comes from the range of likely  surface magnetic field strengths and galactic accelerations.

\cite{Strader2019} reported the masses of 10 redbacks and redback candidates, finding an average mass for redback neutron stars of $1.78 \pm 0.09$\,\msun\ with a dispersion of $\sigma = 0.21 \pm 0.09$\,\msun. We therefore examine the viability of a 1.8\,\msun\ pulsar. A 1.8\,\msun\ pulsar with a 0.09\,\msun\ companion predicts a $\dot{P}_{\mathrm{orb}}$ of $-2.59^{+0.17}_{-0.24} \times 10^{-13}$, consistent with our measured value of $-2.5(3) \times 10^{-13}$. For the lowest reasonable values of magnetic field and galactic acceleration, a 1.9$-$2.0\,\msun\ pulsar is still consistent with our measured value of $\dot P_\mathrm{orb}$. Although this higher mass pulsar cannot be definitively ruled out, it seems unlikely that it could have reached this mass while only being spun up to an 11.56\,ms pulse period.

We therefore cannot make strong constraints on Ter5A's mass; it is most likely in the range of 1.5$-$1.8\,\msun. As a less-recycled redback, it is likely also a less massive one. We conclude that $\dot{P}_\mathrm{orb}$ is dominated by gravitational wave radiation. Our results validate our assumption that $\dot{P}_{\rm RB}$ is close to zero.

To place an upper limit on the contribution of $A_\mathrm{RB}$, we compare our predicted and measured values. If we do indeed have a 1.5\,\msun pulsar with a 0.09\,\msun companion, the contribution of $\dot{P}_\mathrm{RB}$ is no greater in magnitude than $6 \times 10^{-14}$, or $\sim$20\% of the observed orbital period derivative. Accounting for unlikely edge cases in intrinsic spindown rate and galactic acceleration, even if our predicted $\dot P_\mathrm{orb}$ were as low as $-2.0 \times 10^{-13}$, and if the true $\dot P_\mathrm{orb}$ were at the upper limit of our measurement, $-2.8 \times 10^{-13}$, $A_\mathrm{RB}$ cannot account for more than $30\%$ of the observed $\dot P_\mathrm{orb}$. We take this 30\% value as an upper limit, but we do not need any $\dot{P}_{\rm RB}$ to explain our results. 

\begin{figure}
    \centering
    \includegraphics[width=\linewidth]{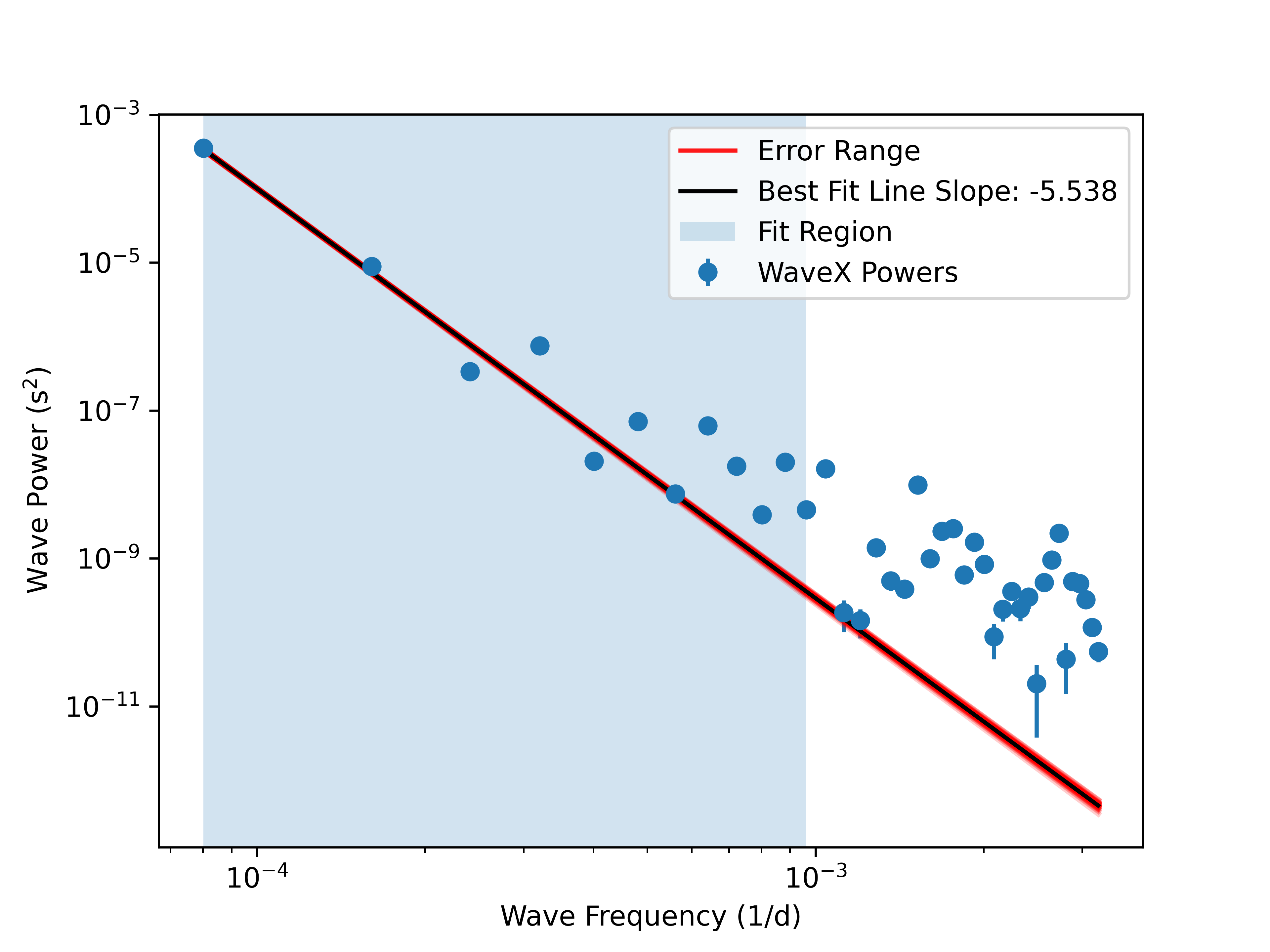}
    \caption{Fourier component analysis of spin residual variations with F0 and F1 subtracted out (see Figure \ref{fig:FreqDerivPlots}, top panel). The error band was estimated by multi-variate gaussian sampling using the fit covariance matrix. The first point and to a lesser extent the second point are potentially underestimations of the true power; we do not adjust for this.}
    \label{fig:spinwaves}
\end{figure}

\begin{figure}
    \centering
    \includegraphics[width=0.5\textwidth]{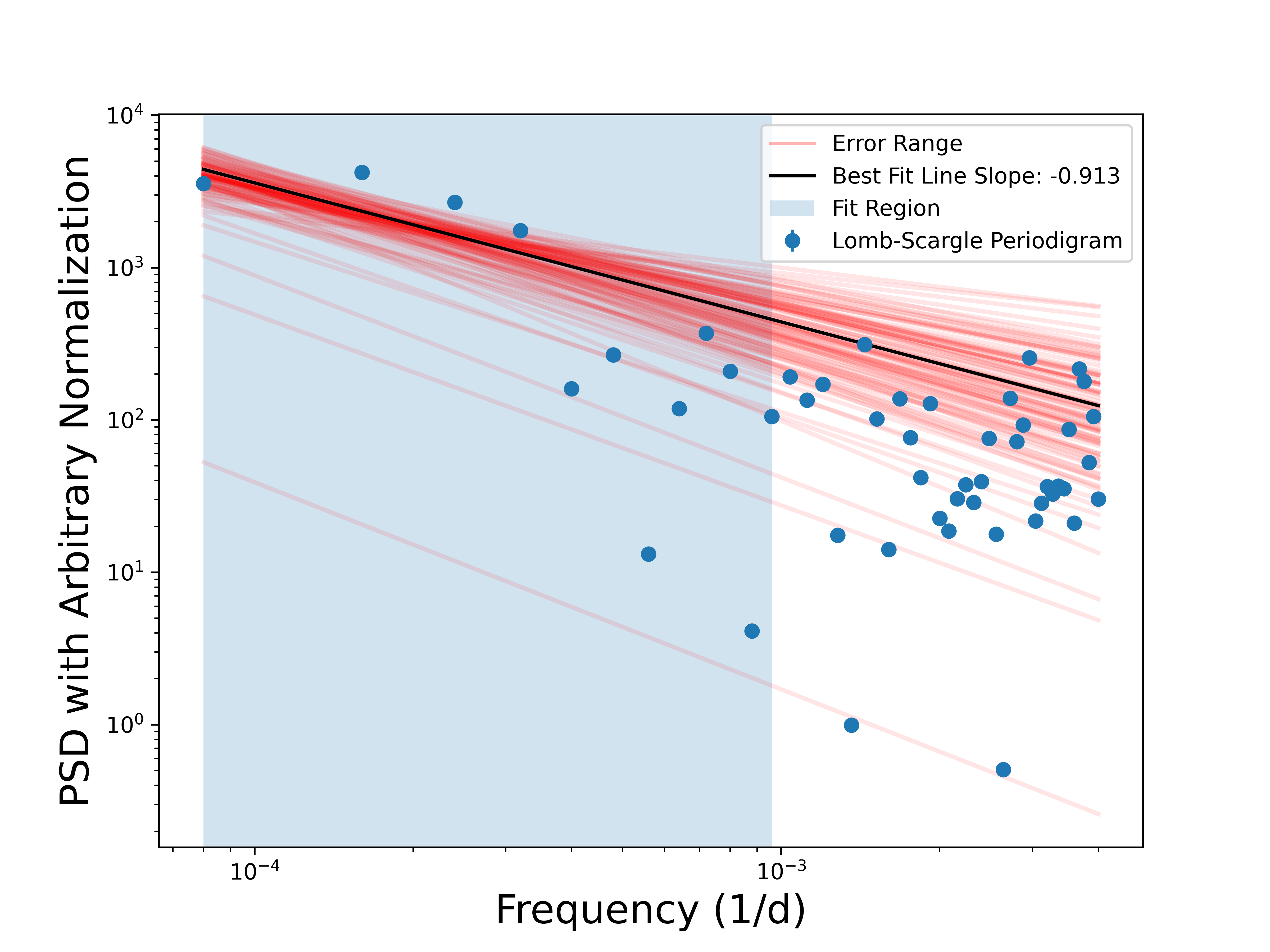}
    \caption{PSD of T0 variations (excluding outliers; see Figure \ref{fig:GPR}) with $\dot P_\mathrm{orb}$ removed. Because $\dot P_\mathrm{orb}$ is quadratic and therefore partially covariant with a single sinusoid, the first bin is likely an underestimate of the true power. We do not adjust our calculations for this underestimation. The steep-spectrum, as well as the oscillations dominating Figure~\ref{fig:GPR}, demonstrate that the signal is dominated by a red-noise process. The error band was estimated by multi-variate gaussian sampling using the fit covariance matrix.}
    \label{fig:PSD}
\end{figure}

\begin{table}[h!]
    \centering
    \begin{tabular}{c@{\hskip -50pt}D{.}{.}{-1}}
        \hline
        Component & \text{Value} \\
        \hline
        $A_\mathrm{spin,obs}$ & -4.62 \times 10^{-10} \,\mathrm{m}\,\mathrm{s}^{-2} \\
        $A_\mathrm{int}$ & 14.00 \times 10^{-10} \,\mathrm{m}\,\mathrm{s}^{-2} \\
        $A_\mathrm{gal}$ & 3.39 \times 10^{-10} \,\mathrm{m}\,\mathrm{s}^{-2} \\
        $A_\mathrm{Shk}$ & 1.41 \times 10^{-10} \,\mathrm{m}\,\mathrm{s}^{-2} \\
        $A_\mathrm{GC}$ & -21.52 \times 10^{-10} \,\mathrm{m}\,\mathrm{s}^{-2} \\
        $A_\mathrm{GR}$ & -85.47 \times 10^{-10}\,\mathrm{m}\,\mathrm{s}^{-2} \\
        $A_\mathrm{orb,obs}$ & -110.9(18.0) \times 10^{-10}\,\mathrm{m}\,\mathrm{s}^{-2}  \\
        \hline
        Predicted $\dot{P}_\mathrm{orb}$ & -2.34^{+0.22}_{-0.17} \times 10^{-13} \\
        Measured $\dot{P}_\mathrm{orb}$ & -2.5(3) \times 10^{-13} \\
        \hline
    \end{tabular}
    \caption{Predicted contribution of different accelerations to orbital period derivative. The listed value of $A_\mathrm{int}$ presumes a magnetic field of $8.0 \times 10^8$\,G, which is the center of the reasonable distribution. The listed value of $A_\mathrm{gal}$ is precise to a factor of two \citep{AGal}. Assuming a distance to the pulsar of $6.6$\,kpc with a $\sim$10\% error and the proper motion of the entire cluster $5.61(7)$\,mas/yr \citep{ClusterPM} yields a value of $A_\mathrm{Shk} = 1.41(15) \times 10^{-10}$\,m\,s$^{-2}$. $A_\mathrm{GC}$ is calculated from these values. The listed value of $A_\mathrm{GR}$ assumes a 1.5\,$M_\odot$ pulsar with a 0.09\,\msun companion. Errors on other components are discussed in depth in section \ref{PBDOT}.}
    \label{tab:acc}
\end{table}

We note that the time for a binary to coalesce due to GR is given by \citep[e.g.,][]{S&T}:
\begin{equation}
    t = \frac{5}{256}\frac{G^3}{c^5}\frac{r^4}{m_1m_2(m_1+m_2)},
\end{equation}
where $r$ is the total separation between the two objects. For Ter5A, if the companion was a compact star that would lose no mass, the coalescence time would be $\sim$250\,Myr. In reality, though, the companion is not compact and more complex binary interactions would occur as the orbit shrinks, dramatically changing this GR-driven evolution.

We also note that we do not expect a measurable value of the contraction of the semimajor axis in timing residuals due to the contraction of the orbit from gravitational wave emission. Over the time period of our observations, $\dot P_\mathrm{orb}$ predicts a change in period that, via Kepler's third law, corresponds to a fractional change of $2.3\times 10^{-8}$. The fractional error of our timing semi-major axis measurement is $6.4\times10^{-6}$, so the change in semimajor axis is significantly smaller than our errors and is currently undetectable via timing.

In theory, the relativistic precession of the system \citep[i.e.,][]{OMDOT}, predicted to be $\sim$20\,$^\circ$/yr, should be easily measurable. However, in practice, the eccentricity is so close to zero that this is not possible --- the precession is entirely covariant with the orbital period. 

\begin{figure*}
    \centering
    \includegraphics[width=\linewidth]{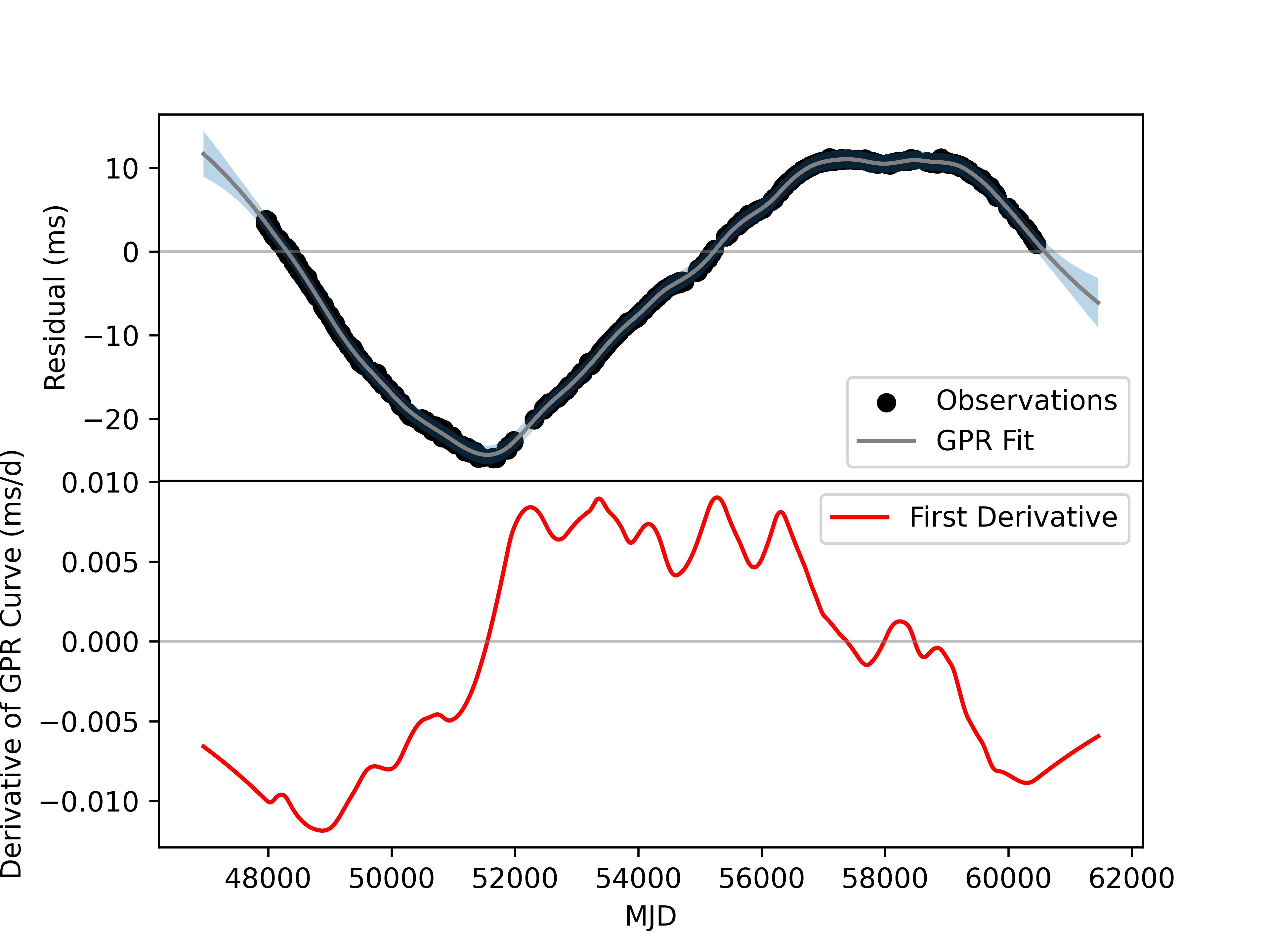}
    \caption{The F0- and F1-subtracted pulse residuals and their GPR fit are plotted in the top panel, and the first derivative of this curve is plotted in the bottom. Note the strong feature around MJD 51500 corresponds to a steep slope in the first derivative plot, which is likely related to a torque on the neutron star.}
    \label{fig:torques}
\end{figure*}

\subsection{Position and Proper Motion} \label{pospm}

Fitting for position over the entire dataset yields a value that is inconsistent with VLA imaging positions and prior timing-based positions. The source of this discrepancy is two sided. Eclipse variability during long integrations can cause systematic positional offsets in VLA imaging \citep[for further discussion of positional uncertainties, see][]{FG2000}. On the other hand, long-term pulsar timing systematics also produce uncertainties and can bias the position measured in the timing analysis. However, fitting for position over just 2-3 years of data at a time produces reasonable positions that are both self-consistent and consistent with the proper motion measured by \textit{Gaia} \citep{ClusterPM}. While we cannot get extremely precise positions with this method, as Figure~\ref{fig:positions} shows, it is possible to achieve sub-arcsecond positions, with the caveat that Ter5A's very low ecliptic latitude ($\beta = -1.4\degree$) means declination measurements via timing are highly imprecise.

Fitting for proper motion over the entire dataset likewise produces unphysical results; again, long-term timing systematics skew the proper motion measurements. Throughout this paper, we often present measurement uncertainties for the last two significant digits in parenthesis after the measurement. Timing over the entire dataset measures a proper motion in the RA direction of $-1.53(11)$\,mas/yr and in the Dec direction of $-24.5(3.2)$\,mas/yr. The dispersion in the proper motion of Terzan 5 stars is $\sim$0.5\,mas/yr \citep{PMdispersion,GAIAEDR3}, so while the RA-proper motion is plausible in comparison with \textit{Gaia}-measured values ($-1.53(11)$\,mas/yr compared to $-1.989(68)$\,mas/yr) \citep{ClusterPM}, the declination-proper motion radically differs ($-24.5(3.2)$\,mas/yr compared to $-5.243(66$)\,mas/yr). The source of this discrepancy is likely a combination of the effects of low ecliptic latitude on declination precision and covariance with other measured parameters. 

The large timing declination proper motion is also clearly unphysical. If Ter5A is indeed a cluster member, its elliptical orbit about the cluster core would never take it on the measured trajectory. It cannot even be sensibly ejected in this direction. For an ejection, we would expect a direction of proper motion pointing radially away from the cluster, but the timing-measured proper motion is almost perpendicular to that. While in principle these results could suggest that Ter5A is not actually a cluster member, this is highly unlikely. The observed positive spin frequency derivative practically guarantees cluster membership. 

However, if we instead fit for the position of the pulsar over just a few years of well-sampled, smoothly behaved data at a time, we get results in the RA-direction that are consistent with \textit{Gaia}'s proper motion measurements, as shown in Figure~\ref{fig:positions}. However, the declination results are too ridden with systematic errors to draw any conclusions.

We compare our timing positions to VLA positions in Table \ref{tab:positions} and Figure \ref{fig:positions}. The details of the first two VLA positions are described in \cite{FG2000} and \cite{Urquhart2020}. The last Ter5A position comes from new (2022) VLA continuum images that have been astrometrically corrected based on pulsar timing positions propagated to the appropriate epoch using \textit{Gaia} proper motions (Urquhart et al, in prep). The images were shifted by matching to 20 timed pulsar positions, though Ter5A was excluded from the frame shift calculation. The new positional uncertainties come from the quadrature sum of the rms associated with the pulsar frame shift and the thermal fluctuations in the continuum position fitting. With Ter5A being such a bright source, we are dominated by the pulsar frame shift uncertainties.

\begin{figure*}
    \centering
    \includegraphics[width=\linewidth]{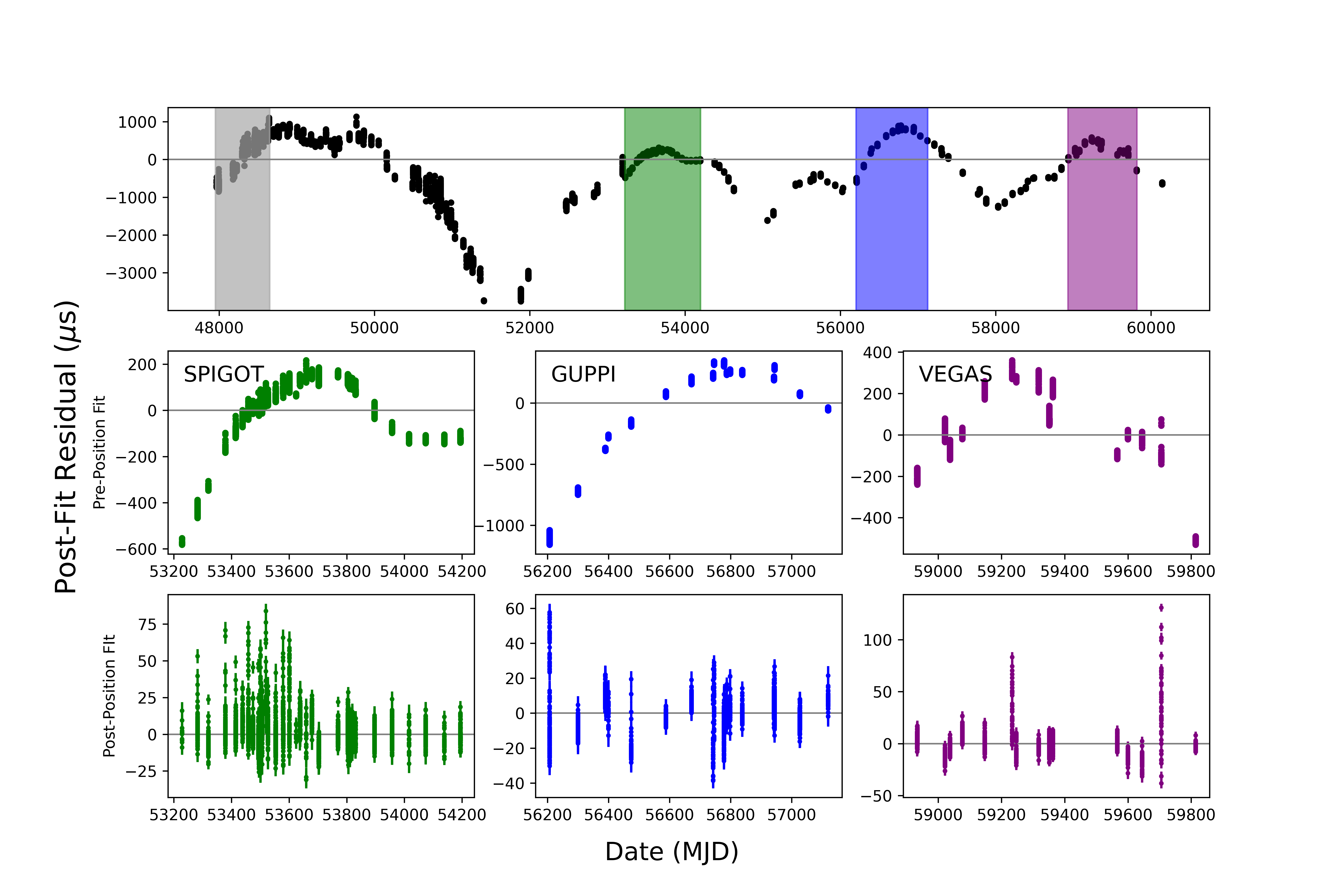}
    \caption{Fitting over a few years of well-sampled data at a time results in positions that are reasonably consistent with \textit{Gaia} proper motion measurements of Terzan~5 as a whole (see Figure~\ref{fig:positions}). The top panel shows our long-term timing solution shown in the third panel of Figure~\ref{fig:FreqDerivPlots}, exempting JB data. The green, blue, and purple highlighted regions correspond to the data ranges used to fit for position. They correspond to different GBT backends and are labeled as such. The second row of plots are the highlighted TOAs in the long-term model. The bottom row are those same TOAs fit for frequency and frequency derivatives 1-4, DM, and position. We use only high-precision GBT data for these fits. These positions are reflected in Table~\ref{tab:positions} and Figure~\ref{fig:positions}. None of the fitted parameters are covariant $>$ 0.8 with position. The gray band corresponds to TOAs used in \cite{Nice1992} to measure position, which is also tabulated in Table~\ref{tab:positions} and plotted in Figure~\ref{fig:positions}.}
    \label{fig:zoom}
\end{figure*}

As a result, we use the SPIGOT timing position and the proper motion measured by \textit{Gaia} \citep{ClusterPM} for our long-term timing solution. These are the parameters that produced the residuals in Figure~\ref{fig:FreqDerivPlots}.

\subsection{Comparison with X-Ray Data} \label{xray}
Ter5A has an associated variable X-ray source, and it is one of the hardest-spectrum and X-ray brightest MSPs in Terzan 5 \citep{XRayID,Bahramian2020}. This source might indicate intermittent accretion. Notably, Ter5A has been observed to have high- and low- X-ray emission states, with approximately 40\% of its X-ray counts measured during two comparatively short observations where it was $\sim$9 times brighter than during other observations \citep{XRayID}. In the ``high"-flux state, \cite{XRayID} reports an unabsorbed 0.5-8\,keV X-ray luminosity of $L_{X} = 8.6(2.1) \times 10^{31}$\,ergs/s, assuming a distance of 5.9\,kpc to the cluster. In the ``low"-flux, or quiescent state, this luminosity is $L_X < 1.3 \times 10^{31}$\,ergs/s with a photon index of $1.5 \pm 2.0$ \citep{XRayID}, which places it in the redback-black widow overlap region in $L_X$--photon index space (see Figure 10 of \cite{Urquhart2020}). The quiescent $L_X$ is lower than that of a typical redback, probably due to a lower-than-typical spindown luminosity. 

\cite{XRayID} suggest two possible interpretations of the data: 1) these luminosity spikes are short-lived burst or flare activity indicating two transitions from a radio pulsar state to an accretion disk-dominated state (as indicated by the blue lines in the bottom panel of Fig~\ref{fig:FreqDerivPlots}), or 2) flaring of the companion star. The spectral and variability properties suggest intrabinary shock radiation \citep{XRayID}. Transitional millsecond pulsars (tMSPs) have been observed to flare similarly \citep[e.g.,][]{Bogdanov_Archibald}. This flaring could be consistent with magnetic activity from the companion's surface \citep[e.g.,][]{Cho_Bogdanov}.

We do not have contemporaneous radio observations during the flared activity. However, a comparison of Ter5A's X-ray quiet days (in which emission is at Chandra's detection threshold and not much higher) and its flared days with spin frequency residuals and orbital variations is presented in the bottom-most panel of Figure~\ref{fig:FreqDerivPlots}. There is no apparent correlation between X-ray activity and changes in either the spin or orbital behavior of the pulsar.

We detect Ter5A in the radio almost every time we observe it. Though there are gaps in our observations, we see very little evidence that Ter5A is radio-silent for any extended period of time and therefore very little evidence that Ter5A spends very much time actively accreting. Even if it were radio-silent for extended periods of time, that is not in and of itself a smoking gun for transitional behavior. For example, J1653$-$0158 is a 1.97\,ms binary pulsar in a 75-minute orbit that has only been detected in gamma-rays despite 26 radio searches on 8 different observatories over 11 years across all orbital phases and several wavelengths \citep{Nieder2020}. However, J1653$-$0158's radio non-detections are attributed to it being eclipsed by ablated material the vast majority of the time. Radio disappearances could simply be due to excess gas in the system and not prolonged accretion-driven X-ray emission.

Ter5A's behavior is also in stark contrast with the known redback and transitional MSP M28I. M28I also experiences X-ray flux enhancements of 1-2 orders of magnitude when it switches from pulsar- to disk- state \citep{M28I_Vurgun}. However, in Chandra data, M28I was detected as a pulsar three times from July to September of 2002, as a disk-accreting X-Ray Binary (XRB) on two observations four days apart in August of 2008, and as a pulsar again on three observations from May to November of 2015 \citep{M28I_Vurgun}. While data are sparse, odds are low that M28I was by chance observed on two consecutive observations as an XRB if its presence in that state is rare. Furthermore, \cite{Papitto_M28I} reports that M28I's behavior is consistent with months-long X-ray outbursts. We see no evidence for months-long state changes in Ter5A. M28I is likely stable for much longer periods of time than Ter5A is. We conclude that Ter5A is likely not a transitional MSP. If it is switching between accreting and non-accreting states, these switches are rare and short in duration.

\subsection{Energetics} \label{energetics}

The Applegate model predicts orbital variability due to deformations in the companion star from magnetic activity. It could potentially explain the observed orbital variations if those variations have a clear periodicity.

We investigate if the Applegate model is energetically viable. We assume that the pulsar is 1.6\,\msun and that the companion star is 0.1\,\msun and approximately the radius of its Roche lobe, using \cite{Nice1990}'s value of 0.2\,$R_{\odot}$. Consulting a $P$-$\dot{P}$ diagram, we note the intrinsic spindown of this pulsar is likely of order $\dot{P} \sim 10^{-19}-10^{-20}$. We assume $\dot{P} = 10^{-19}$ for the following calculations. The energy radiated by the pulsar, or its spindown luminosity, is given by $L_\mathrm{spin} = 4\pi^2I\dot{P}P^{-3}$, where $I$ is taken to be $10^{45}$\,g cm$^2$. Assuming isotropic emission, the companion captures an energy of $L_\mathrm{capt,iso} = 3.3 \times 10^{31}$\,ergs/s of this spindown energy, or 0.008\,L$_{\odot}$. This energy blows mass off the companion at a maximum rate given by
\begin{equation}
    \dot{M}_c = \frac{L_\mathrm{capt,iso} \times R_l}{GM_c},
\end{equation}
which predicts a value of $1.5 \times 10^{-12}$\,\msun/yr. This agrees with predictions made by \cite{Shaham1995} and results in an evaporation time much longer than a Hubble time.

Accounting for the possibility of anisotropic emission, if the companion were to absorb half of the pulsar's total spindown energy (i.e. if a jet was pointed at the companion), the available energy due to spindown radiation would be $L_\mathrm{capt,an} = 1.3 \times 10^{33}$\,ergs/s, or 0.3\,L$_{\odot}$. In the other direction, if the companion were to absorb much less than $L_{\mathrm{capt,iso}}$ (i.e. by an anisotropic model in which most of the energy misses the companion), the captured luminosity would remain of order $10^{30}-10^{31}$\,ergs/s, or $10^{-4}-10^{-3}$\,L$_\odot$.

The energy available due to tidal heating, denoted $L_T$ for tidal luminosity, is given by 
\begin{equation}
    L_T = 4 \times 10^{32} \left(\frac{10^{11}\,\mathrm{cm}}{a}\right)^2 \frac{M_c}{1.6\,M_{\odot}} \frac{10^8\,\mathrm{yr}}{t_M} \, \mathrm{ergs/s}
\end{equation}
according to Equation 28 of \cite{ApplegateShaham1994}, where $t_M$ is the evaporation time in years of the companion and $a$ is the separation. This yields $L_T = 1.0 \times 10^{29}$\,ergs/s, or $2.5 \times 10^{-5}$\,L$_\odot$.

The Applegate model predicts a change in transported angular momentum $\Delta{J}$ given by Equation 27 in \cite{Applegate}:
\begin{equation}
    \Delta{J} = \frac{1}{6\pi}\frac{GM_c^2a^2}{R_c^3}\Delta{P_\mathrm{orb}},
\end{equation}
where $R_c$ is the radius of the companion, here taken to be the Roche lobe radius. The associated energy is given by 
\begin{equation}
    \Delta{E} = \Omega_\mathrm{diff}\Delta{J},
    \label{eqn:rewrite}
\end{equation}
where $\Omega_\mathrm{diff}$ is the differential rotation between layers of the star. This is not measurable, but since the Applegate model requires $\Omega_\mathrm{diff}$ to be at least $\sim$0.01 of the total orbital angular velocity $\Omega$ of the system \citep{Applegate}, and measured constraints suggest that $\Omega_\mathrm{diff}$ can span from $\sim$0.01 to $\sim$0.1 \citep[e.g.,][]{Omegadiff}, we will take the generous assumption that $\Omega_\mathrm{diff}=0.1\Omega$ to test the model's energetic viability. Observations of stars suggest that Noting that $\Delta{P}_\mathrm{orb}/\Delta{t} \sim \dot{P}_\mathrm{orb}$ for $\Delta t$ the length of the dataset, Eqn~\ref{eqn:rewrite} can be rewritten to express the luminosity predicted by the Applegate model $\Delta L_A$:
\begin{equation}
\begin{split}
   \Delta L_A = & \; 13 \times 10^{33} \left(\frac{M_c}{\mathrm{M}_\odot}\right)^2 \left(\frac{10^{10}\, \mathrm{cm}}{R_c}\right)^{3} \\
   & \times \left(\frac{a}{10^{10}\,\mathrm{cm}}\right)^{2} \left(\frac{\dot{P}_\mathrm{orb}}{10^{-9}}\right) \mathrm{ergs/s}.
\end{split}
\end{equation}
This yields $\Delta L_A = 4.6 \times 10^{29}$\,ergs/s, or $1.1 \times 10^{-4}\,\mathrm{L}_\odot$. We conclude that the Applegate model is energetically viable - there is sufficient spindown energy ($\sim 0.01\,\mathrm{L}_\odot)$, regardless of isotropic or anisotropic emission.

While energetically viable, we observe no periodicity in the T0 measurements in Figure~\ref{fig:GPR} to suggest that this process is taking place on a timescale even of a few decades. However, it is certainly possible that some other form of energy exchange is occurring between the orbit and the companion.

\subsection{Torques} \label{torques}

If any of the ablated material falls back on the pulsar, i.e. like a tMSP, or even approaches it \citep[see][]{Nice2000Proceedings}, it should torque the pulsar and affect its spin as the second derivative of phase $\phi$:
\begin{equation}
    \tau = I\frac{d^2\phi}{dt^2}.
\end{equation}
This quantity is related to (but not equivalent to) the second derivative of the timing residuals. From Figure \ref{fig:torques}, it's apparent that the slope of the first derivative is high around the odd feature at about MJD 51500, so that feature does indeed correspond to some type of torque. It is unclear whether this torque is intrinsic (such as internal seismic activity) or extrinsic (such as accretion or the effects of a nearby star), but similar sharp features also appear later in the data. 

These sharp turnarounds are inconsistent with glitches, as the pulses after the turnarounds are showing up later, not earlier, as would be expected for the increase in spin frequency following a glitch. These events all show the pulsar slowing down in its rotation. One possible explanation is that gas interacting with the magnetosphere is hindering Ter5A's rotation, but the mechanism driving these spin features is very unclear.

\subsection{Known Systematics} \label{knownsystematics}

There remain contaminants in the data. In order to limit the number of TOAs and focus on phase connection, much of the radio frequency information available in the original data has been discarded. For an example and analysis of how DM can vary over the course of a single Ter5A observation, see Figure 10 of \cite{Bilous2019BSPs}. This figure is an analysis of the first observation in the GUPPI panel in Figure~\ref{fig:zoom}, and shows a DM change of $\sim$0.6\,pc\,cm$^{-3}$ in a systematic manner over the course of the observation. This is an example of a day in which DM variations -- which we do not account for in our study -- prohibit precise measurement of T0. Similar DM-related effects are likely present during most observations, but with smaller amplitudes.

Of the nine redback pulsars with published long-term timing solutions \citep{Thongmeearkom2024,Corcoran2024}, Ter5A has by far the most timing noise. Six redbacks show effectively flat timing residuals when fit out to the second frequency derivative, and two show flat residuals when fit to the fourth frequency derivative \citep{Thongmeearkom2024,Corcoran2024}, while Ter5A shows systematics at the $\sim$1\,ms level even when fit to the fifth frequency derivative.

Unmodeled DM variations, either short- or long-term, cannot account for even a small percentage of the remaining timing noise. If we assume most of our data has a central frequency of 1800\,MHz, a DM variation of 12\,pc\,cm$^{-3}$ would be required to create the $\pm$15\,ms of timing variation seen in Figure~\ref{fig:FreqDerivPlots}. Inspection of the radio frequency dependence of each observation ensured that the DM was stable to within $\sim$0.2\,pc\,cm$^{-3}$ on all but a couple of days. We can rule out DM as a significant contaminant in the remaining timing noise.

This timing noise could be caused by torques on the pulsar due to infalling matter, or by other dynamical effects due to the dense cluster environment. While it is surprising that the cluster could play such a role given that Ter5A is located far from the cluster core, the pulsar's observed negative spin period derivative indicates that it has to. Since cluster effects dominate the first period derivative measurement, they must contribute to higher order derivative measurements as well. We do not see these effects reflected in the orbit because the pulsar's spin is much more sensitive than the orbit (see Section~\ref{sfd} for discussion).

\section{Conclusions \& Future Work} \label{concs}

We have determined a 34-year duration timing solution for the eclipsing millisecond pulsar Terzan 5A, the longest timing solution for any redback system. We successfully used novel timing methods such as iterative parameter-fitting, employing a continuous binary piecewise model, and using precise binary values to remove the orbital motion of the pulsar and effectively isolate it.

There are strong systematic timing residuals, but they don't seem to directly correspond to measured orbital variations. The amount of timing noise in this system is far greater than that of standard millisecond pulsar systems and complicates or inhibits many traditional pulsar timing methods, including precision measurements of position and proper motion. This timing noise could be caused by torques on the pulsar due to infalling matter, or by other dynamical effects due to the dense cluster environment, even though Ter5A is on the outskirts of the cluster. While we successfully phase-connected 34 years of observations, this is too complex a system to to explain all of the residuals in our data with a simple physical model. Systematic effects remain at the 1\,ms level.

We measured the orbital decay of the system over time and found it consistent with GR predictions for a companion of mass 0.1\,\msun, a highly-inclined orbit, and a pulsar of 1.5 to 1.8\,\msun. Though our measurements are not inconsistent with a higher-mass pulsar for very low effective spin and galactic accelerations, it is unlikely that Ter5A achieved such a mass given its relatively slow spin period. General relativity seems to be the dominant effect in the measured orbital period derivative, which constrains the magnitude of tidal effects in this system to be at most 30\% of the observed acceleration, and likely much less.

We could not use long-term timing to determine precise positions or proper motions of the system because over long-term timing baselines timing noise strongly contaminates the timing signals of these parameters. However, measuring position via timing over few-year periods yielded reasonable position measurements consistent with a measured \textit{Gaia} proper motion for Terzan 5.

The power spectral density of Ter5A's orbital variations is consistent with those of other redback systems, and is roughly power-law in nature, although with a shallower spectrum than some of the other systems. The variations in orbit and sometimes strong changes in spin as measured by pulsar timing do not seem to correlate with measured X-ray emission.

Possible directions for future work include measuring time-variable DMs over the course of all observations, providing insight into Ter5A's environment. Another is to use our model and all TOAs, including those affected by the eclipse, to examine in greater depth the eclipses themselves. Furthermore, the torque on the pulsar can be calculated and compared to wind-based accretion models as well as the precession of the system predicted in \citet{Nice2000Proceedings}. Investigating the systematics in the spin frequency derivative residuals could reveal more about the source of the timing noise.  Are they from e.g.~the companion, internal neutron star processes, cluster dynamics, or infalling gas?

\begin{acknowledgements}
 The National Radio Astronomy Observatory and Green Bank Observatory are facilities of the U.S. National Science Foundation operated under cooperative agreement by Associated Universities, Inc. Murriyang, the Parkes radio telescope, is part of the Australia Telescope National Facility (\url{https://ror.org/05qajvd42}) which is funded by the Australian Government for operation as a National Facility managed by CSIRO. We acknowledge the Wiradjuri people as the Traditional Owners of the Observatory site. SMR is a CIFAR Fellow and is supported by the NSF Physics Frontiers Center award 2020265.  DJN is also supported by NSF award 2020265.
JS acknowledges support from NASA grant 80NSSC21K0628, NSF grant AST-2205550, and the Packard Foundation.
\end{acknowledgements}

\bibliographystyle{aasjournal}
{\scriptsize
\bibliography{main.bib}}

\end{document}